\documentclass[%
reprint,
nofootinbib,
amsmath,amssymb,
aps,
]{revtex4-2}

\usepackage{siunitx}
\usepackage{graphicx}
\usepackage{dcolumn}
\usepackage{bm}

\usepackage{color} 
\usepackage{hyperref} 

\begin{document}
	
	\preprint{APS/123-QED}
	
	\title{Super Bound States in the Continuum on Photonic Flatband: Concept, Experimental Realization, and Optical Trapping Demonstration}
	
	\author{Ngoc Duc Le$^{1,2}$} 
	\author{Paul Bouteyre$^{1,3}$} \thanks{N.D. Le and P. Bouteyre contributed equally to this work as first authors.}
	\author{Ali Kheir-Aldine$^1$}
	\author{Florian Dubois$^{1,4}$}  
	\author{S\'{e}bastien Cueff$^1$}
	\author{Lotfi Berguiga$^1$}
	\author{Xavier Letartre$^1$}  
	\author{Pierre Viktorovitch$^1$} 
	\author{Taha Benyattou$^1$} 
	\author{Hai Son Nguyen$^{1,5}$}     
	\email{hai-son.nguyen@ec-lyon.fr} 
	\affiliation{$^1$ Univ Lyon, ECL, INSA Lyon, CNRS, UCBL, CPE Lyon, INL UMR 5270, 69130 \'{E}cully, France}  
	\affiliation{$^2$ Universit\'{e} Paris-Saclay, CNRS, CEA, Institut de Physique Th\'{e}orique, 91191 Gif-sur-Yvette, France} 
	\affiliation{$^3$ Department of Physics and Astronomy, University of Sheffield, S3 7RH, Sheffield, UK}
	\affiliation{$^4$ Silicon Austria Labs GmbH (SAL), 9524 Villach, Austria}
	\affiliation{$^5$ Institut Universitaire de France (IUF), 75231 Paris, France}

	\date{\today}

	\begin{abstract}
		In this work, we theoretically propose and experimentally demonstrate the formation of a super bound state in a continuum (BIC) on a photonic crystal flat band. This unique state simultaneously exhibits an enhanced quality factor and near-zero group velocity across an extended region of the Brillouin zone. It is achieved at the topological transition when a symmetry-protected BIC pinned at $k=0$ merges with two Friedrich-Wintgen quasi-BICs, which arise from the destructive interference between lossy photonic modes of opposite symmetries. As a proof-of-concept, we employ the ultraflat super BIC to demonstrate three-dimensional optical trapping of individual particles. Our findings present a novel approach to engineering both the real and imaginary components of photonic states on a subwavelength scale for innovative optoelectronic devices.
	\end{abstract}
	
	\maketitle

	\emph{Introduction -} 
	In recent years, the photonic platform has advanced as a fertile research area for exploring non-Hermitian physics \cite{Guo2009,Ruter2010},  where the interaction between the real (i.e., oscillation frequency) and imaginary (i.e., losses) components of photonic resonances can lead to unique phenomena that are absent in Hermitian systems. 
	One archetypal object of non-Hermitian photonics is the \textit{Bound states in the continuum} (BICs) \cite{Hsu2016}, that are perfectly confined modes despite lying in the continuum of radiating waves. 
	These unique states are prevented from radiating due to either a symmetry mismatch with radiating waves in the case of  \textit{symmetry-protected BICs}, or through loss cancellation via destructive interference for \textit{Friedrich-Wintgen BICs} \cite{FriedrichH1985}. 
	In photonic lattice, each BIC is pinned at a singularity of far-field polarization vortex and possesses a \textit{topological charge} determined by the corresponding winding number \cite{Zhen2014}. 
	The topological nature of BICs has been explored by various research groups through numerous theoretical propositions \cite{Zhen2014,Yoda2020,Kang2021,Kang2022,Jiang2023} and experimental demonstrations \cite{Zhang2018,Doeleman2018,Jin2019,Yin2020,Hwang2021,Liu2019,Zhang2022,Chen2023}. 
	However, most of these works only focus on manipulating the imaginary part of BIC modes, such as splitting BICs into lower-order BICs and chiral photonic states with circular polarization \cite{Yoda2020,Liu2019,Zhang2022,Chen2023}, or merging BICs to achieve \textit{super BICs} characterized by a remarkably high quality factor over an extended region of the Brillouin zone \cite{Jin2019,Hwang2021,Kang2021,Letartre2022,Jiang2023}. 
	
	In this Letter, we investigate the complex energy-momentum dispersion of photonic band structures when merging topological charges originated from a symmetry-protected BIC (sym-BIC) at $\Gamma$-point and two Friedrich-Wintgen quasi-BICs (FW-qBICs). 
	Both BICs are located at photonic band-edges of opposite curvature, and their merging gives rise to a super BIC sitting on an ultra-flat band of zero-curvature and infinite effective mass. 
	All of the results obtained from numerical simulations are in good agreement with  an analytical model based on mode coupling theory that nicely reproduces the real part of the band dispersion, the photonic quality factor and the corresponding far-field polarization texture. 
	Transferring these concepts to real materials with technological constraints, the formation of flatband resulting from merging topological charges carried by BICs is experimentally demonstrated. 
	Furthermore, we leverage this super BIC state to achieve the first experimental demonstration of optical trapping using BIC states. 
	Our findings lay the groundwork for engineering photonic modes with robust quality factors and ultra-high density of states, paving the way for innovative applications in lasing, particle trapping and sensing.

	\begin{figure}
		\centering
		\advance \leftskip -2mm  
		\includegraphics[width=0.48 \textwidth]{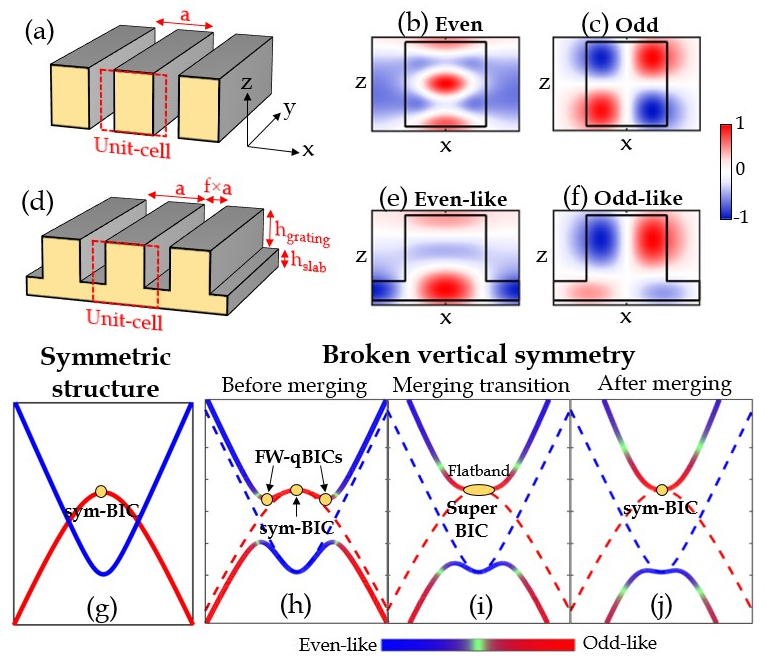}
		\caption{(a) Sketch of 1D photonic crystal slab having vertical symmetry. 
			Distribution of the electric field of (b) the even symmetric and (c) the odd antisymmetric modes of the TE-polarization in the structure (a). 
			(d) Sketch of 1D photonic crystal ``comb" structure. 
			Distribution of the electric field of (e) the even-like symmetric and (f) the odd-like antisymmetric modes of the TE-polarization in the structure (d). 
			(g-j) The dispersion relations for $k_y = 0$ of: 
			(g) Uncoupled even (blue) and odd (red) modes. 
			(h) The coupling between odd-like and even-like modes gives three BICs on a multivalley band. 
			(i) The merging of the three BICs results in a super BIC on a flatband. 
			(j) As the coupling coefficient $\alpha$ gets stronger, the band becomes parabolic. 
		}
		\label{fig:Concept}
	\end{figure}
	
	\emph{Concept -} 
	We start by considering a 1D photonic crystal slab having vertical symmetry $(-z \rightarrow z)$ and in-plane symmetry $(-x \rightarrow x)$ (Fig.~\ref{fig:Concept}a). 
	The structure has period $a = $\SI{330}{\nano \meter}, total thickness $h = $\SI{350}{\nano \meter}, and refractive index $n_{PC} = 3.15$. 
	We define the filling factor $f$ as the proportion of one unit cell filled with air. 
	To avoid any confusion, in this manuscript, we use the terms ``odd/even" to describe the parity of the vertical symmetry, and the terms ``symmetric/antisymmetric" for the parity of the in-plane symmetry. 
	Figs.~\ref{fig:Concept}b and c depict the profiles of the electric fields at the $\Gamma$-point of the two lowest TE-polarized modes, showing that the fundamental mode is even symmetric and the first excited mode is odd antisymmetric. 
	Due to symmetry mismatch, these two modes are unable to couple with each other.
	Moreover, due to the symmetric parity, the fundamental mode is able to couple to the plane waves of the radiative continuum and becomes a leaky mode. 
	By contrast, due to the antisymmetric parity, the first excited mode cannot couple to the radiative continuum, and forms a sym-BIC at the $\Gamma$-point (Fig.~\ref{fig:Concept}g).   
	
	Now, we break the vertical symmetry by implementing a thin unpatterned slab of thickness $h_{slab} =$\SI{80}{\nano \meter} beneath the corrugated grating to create a ``comb" structure. 
	The height of the grating is $h_{grating} =$ \SI{270}{\nano \meter}.
	The fundamental and first excited modes are no longer even and odd, become even-like symmetric (Fig.~\ref{fig:Concept}e) and odd-like antisymmetric (Fig.~\ref{fig:Concept}f), respectively. 
	At the $\Gamma$-point, the antisymmetric nature of the odd-like mode prevents it to couple to neither the even-like mode nor the radiative continuum.
	Hence, it still exhibits a sym-BIC at the $\Gamma$-point.
	Out of the $\Gamma$-point, the in-plane symmetry is broken, allowing the odd-like and even-like modes to couple to each other. 
	The coupling constant becomes stronger when we go further away from the $\Gamma$-point, so we can approximate the coupling strength as $U = \alpha |k_x|$. 
	Because this coupling mechanism involves the breaking of both the vertical and in-plane symmetries, we call it \textit{symmetry-breaking coupling}. 
	This coupling mechanism is described by the Hamiltonian:  
	\begin{equation}
		H_{comb} = \begin{bmatrix}
			\omega_o & U 
			\\
			U & \omega_e 
		\end{bmatrix}
		+ i \begin{bmatrix}
			\gamma_o & \Gamma e^{-i\phi} 
			\\
			\Gamma e^{i\phi} & \gamma_e 
		\end{bmatrix} 
		\label{eq:CouplingHamiltonian} 
	\end{equation}
	
	Here $\omega_{e(o)} (k_x,k_y)$ and $\gamma_{e(o)} (k_x,k_y)$ are the real and imaginary parts of the uncoupled even (odd) modes. 
	The anti-diagonal term $\Gamma$ corresponds to the coupling rate between the two modes via the radiative channels. 
	The phase-shift $\phi$ is the relative dephasing when they radiate into the continuum. 
	We focus on the case $k_y = 0$. 
	The substantial hybridization between the even-like and odd-like modes opens a gap at the anticrossing points of the dispersion curves, resulting in two bands of multivalley shape (Fig.~\ref{fig:Concept}h).  
	The loss exchange through the mode coupling mechanism reaches its peak when the Friedrich-Wintgen condition for destructive interference is satisfied: $ \omega_o - \omega_e = \text{sign} (Z) \sqrt{\dfrac{W+V}{W-V}} (\gamma_o - \gamma_e)$ with $W = \sqrt{U^4 + 2U^2\Gamma^2 \cos 2 \phi + \Gamma^4}$, $V = U^2 - \Gamma^2$ and $Z = U\Gamma \cos \phi$.  
	Therefore, one mode takes almost radiative loss and becomes a leaky mode, while the other mode experiences a minimal loss and becomes a FW-qBIC localized near the anticrossing point.   
	
	As the coupling coefficient $\alpha$ gets stronger, the symmetry-breaking coupling intensifies through the whole reciprocal space, drawing the anticrossing points followed by the FW-qBICs towards the $\Gamma$-point. 
	The coupling coefficient $\alpha$ can be controlled by varying the filling factor $f$\,\cite{supp}. 
	This motion flattens the upper band.
	Therefore, a ultraflat band emerges when the BICs merge into a super BIC \cite{Jin2019,Hwang2021}.  
	This super BIC has a large extension over a broad range of the dispersion curve.
	In other words, it is robust against oblique angles. 
	Our work stands out from previous studies, which exclusively concentrated on the quality factor enhancement \cite{Jin2019,Hwang2021,Kang2021} or photonic flat bands focusing on density of states enhancement \cite{Nguyen2018,Cueff2019,Tang2021,Dong2021,Nguyen2022,Fayard:22, Letartre2022, Yang2023}. 
	We introduce the concept of a \emph{super bound state in the continuum on a flat band}, referred to as an \textit{ultraflat super BIC}, which shows a dual effect: 
	i) Quality factor enhancement due to the super BIC nature, 
	and ii) Density of states enhancement resulting from the flat band characteristic. 
	
	\begin{figure}[h] 
		\centering
		\includegraphics[width=0.48 \textwidth]{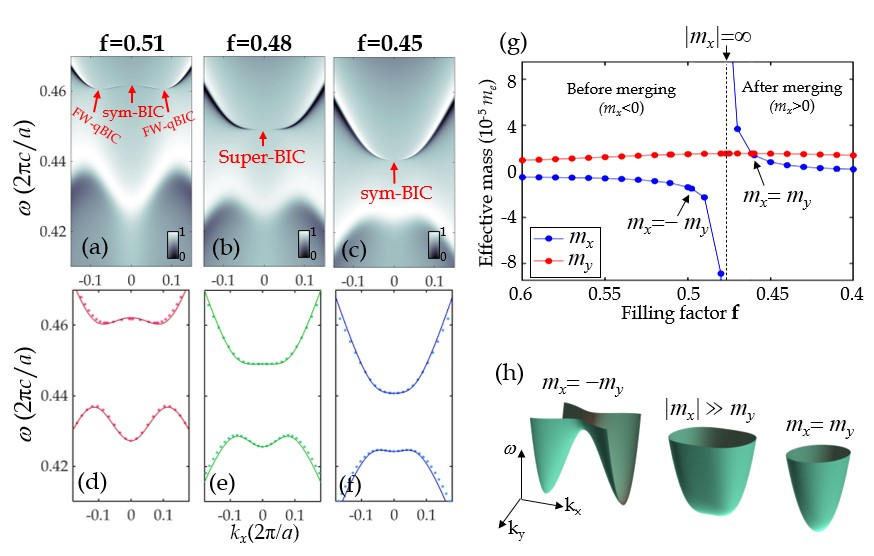}
		\caption{(a-c) Numerical reflectivity spectra of the ``comb" structure of filling factors $f=0.51,0.48,0.45$, obtained by RCWA simulations ($k_y = 0$).
			(d-f) Comparison between the dispersion curves obtained from the RCWA simulations (dots) and from the analytical model \eqref{eq:CouplingHamiltonian} (solid line) for $k_y = 0$.
			(g) Effective masses $m_x$ and $m_y$ at the $\Gamma$ point of the upper band as function of the filling factor $f$ (RCWA simulations).  
			(h) Dispersion surfaces correspond to the case of saddle surface ($f=0.50$), ultraflat band  ($f=0.48$) and paraboloid surface ($f=0.46$) (calculated using the COMSOL software).} 
		\label{fig:RealPart} 
	\end{figure}
	
	\emph{Engineering the real part dispersion-} Figs.~\ref{fig:RealPart}a-c display the reflectivity spectra of the ``comb" structure for $k_y = 0$ for varying filling factor $f$, obtained using the numerical RCWA method \cite{Moharam1995}. 
	These results reveal that before the merging transition ($f=0.51$, Fig.~\ref{fig:RealPart}a), a sym-BIC and two FW-qBICs coexist in the upper band, characterized by a multi-valley-shaped dispersion curve. 
	At $f=0.48$, the three BICs merge, forming a super BIC with a ultra-sharp linewidth spanning a large portion of the Brillouin zone (Fig.~\ref{fig:RealPart}b). 
	After the merging transition ($f=0.45$, Fig.~\ref{fig:RealPart}c), only the sym-BIC remains. 
	The RCWA results are accurately replicated by the analytical model based on the Friedrich-Wintgen coupling mechanism in \eqref{eq:CouplingHamiltonian}\,\cite{supp}, with the results depicted by the solid lines in Figs.~\ref{fig:RealPart}d-f. 
	Concerning the dispersion characteristics along the $k_y$-direction, they exhibit a simple parabolic dispersion of free particles with nearly constant positive curvature for all values of the filling factor $f$, since the corrugation is only along the $x$-axis.
	
	To summarize the dispersion properties along both $k_x$ and $k_y$, Fig.~\ref{fig:RealPart}g displays the photon effective masses $m_x$ and $m_y$ at the $\Gamma$ point of the upper band. 
	As previously mentioned, $m_y$ remains almost constant with positive value as the filling factor $f$ changes. 
	Interestingly, $m_x$ is negative before the merging transition, corresponding to the multi-valley dispersion in Fig.~\ref{fig:RealPart}a. 
	Its absolute value grows to infinity at the merging transition, indicating the presence of a ultraflat band, as shown in Fig.~\ref{fig:RealPart}b. 
	After the merging transition, $m_x$ becomes positive, corresponding to the parabolic dispersion with positive curvature in Fig.~\ref{fig:RealPart}c. 
	Notably, $m_x=m_y$ at $f=0.46$, where the dispersion relation is isotropic. 
	Another special case occurs at $m_x=-m_y$ with $f=0.50$. 
	The dispersion surfaces of these configurations, as well as the ultraflat band configuration, are illustrated in Fig.~\ref{fig:RealPart}h, depicting a transformation from a saddle surface to a paraboloid surface. 
	It shows that isotropic dispersion surfaces can arise from such totally anisotropic configurations. 
	These findings demonstrate that the ``comb" structure can provide on-demand curvature in the momentum space.

	\begin{figure}[h] 
		\centering
		\includegraphics[width=0.48 \textwidth]{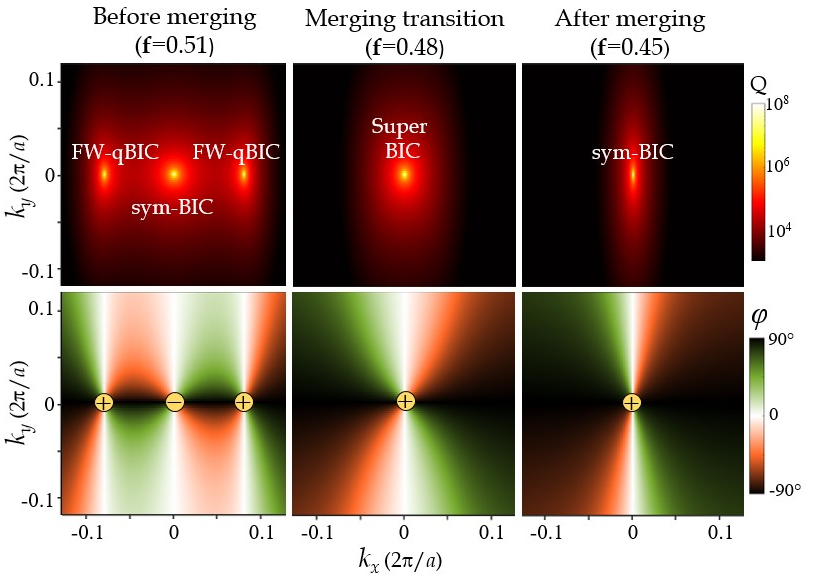}
		\caption{(First row) Quality factor of the upper band as a function of $k_x$ and $k_y$.
			(Second row) Orientation angle $\varphi$ of the far-field polarization of the upper band. These results are obtained by using the analytical model based on the Friedrich-Wintgen formalism \eqref{eq:CouplingHamiltonian}.} 
		\label{fig:MergingTransition}
	\end{figure}
	
	\emph{Engineering the imaginary part and topological charges-}  
	Fig.~\ref{fig:MergingTransition} shows the mapping in momentum space of the quality factor and the orientation angle $\varphi$ of the elliptic polarization of light. 
	The results are obtained by using the analytical model (imaginary part for quality factor calculation, and eigenvectors for polarization texture) and are perfectly reproduced by numerical calculations (see Supplemental Materials \cite{supp}). 
	For the case $f=0.51$, the figure clearly displays a sym-BIC at the $\Gamma$-point and two FW-qBICs on the line $k_y=0$. 
	When $f=0.48$, a super BIC emerges at the $\Gamma$-point, exhibiting a large quality factor extension in momentum space as a result of the merging of the previously mentioned sym-BIC and FW-qBICs. 
	At $f=0.45$, only a sym-BIC remains at the $\Gamma$-point. 
	These results are in perfect agreement with the calculated reflectivity spectrum previously presented in Fig.~\ref{fig:RealPart}. 
	The topological charge carried by each BIC is defined as the winding number of the far-field polarization vortex around this BIC \cite{Zhen2014} and is calculated from the orientation angle $\varphi$ of the elliptic polarization:  
	\begin{equation}
		m = \frac{1}{2 \pi} \oint_C d \mathbf{q} \cdot \nabla_{\mathbf{q}} \varphi    
	\end{equation}
	Our results show that each BIC carries a topological charge which equals to either +1 or -1. 
	The sum of the topological charges is conserved, in agreement with previous works \cite{Jin2019,Hwang2021,Kang2021}. 
	
	\begin{figure}[h]
		\centering
		\includegraphics[width=0.48\textwidth]{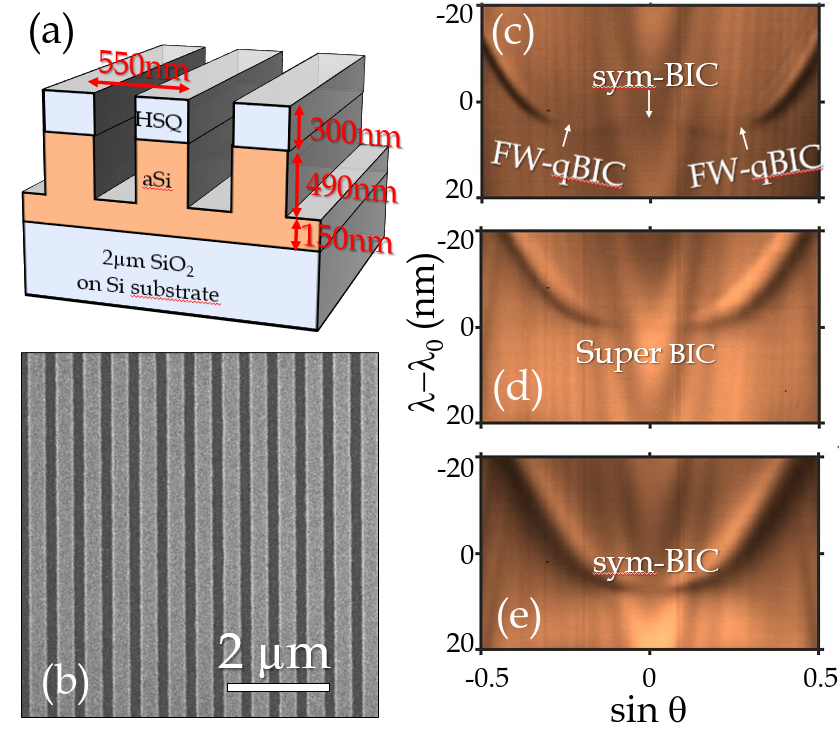} 
		\caption{(a) Sketch of the sample design. (b) SEM image of a grating structure (here $f=0.39$). (c-e) Angle-resolved reflectivity measurements of three structures having three different filling factors $f$ that correspond to (c) before merging, (d) at the merging transition, (e) after the merging transition. To follow the same photonic band, the spectral range of 40 nm is centred at $\lambda_0=$ (a) 1043 nm, (b) 1020 nm, (c) 1007 nm. }
		\label{fig:Experiments} 
	\end{figure}
	
	\emph{Experimental demonstration-} 
	The concept of super BIC on flatband previously developed is adapted for amorphous silicon (a-Si) material (the spectral dependence of a-Si refractive index is shown in \cite{supp}). 
	In our demonstration, the operating wavelength is targeted in the vicinity of \SI{1}{\micro\metre}. 
	Our sample consists of asymmetric a-Si gratings ($h_{grating}=490$ nm, $h_{slab}=100$ nm) on \SI{2}{\micro\metre} of silica on silicon substrate (see Fig.~\ref{fig:Experiments}a). 
	These structures (\SI{80}{\micro\metre}$\times$\SI{80}{\micro\metre}) were first defined in 300 nm thick HSQ resist layer by electron beam lithography, then were transferred subsequently to the a-Si layer via ionic dry etching (more details are reported in Supplemental Information\,\cite{supp}). 
	All structures are of the same period of 550 nm but of different value for the filling factor $f$ that varies from 0.35 to 0.65. 
	Fig.~\ref{fig:Experiments}b represents the scanning electron microscopy (SEM) image of the structure corresponding to $f=0.39$. 
	A white light beam is focused onto the sample via a microscope objective ($\times$50, NA=0.65). 
	The reflectivity is collected via the same microscope in a confocal geometry, and then projected in the Fourier space onto the entrance slit of a spectrometer. 
	Finally, the signal is captured by the sensor of an infrared camera in the spectrometer output\,\cite{supp}.  
	Figs.~\ref{fig:Experiments}c-f present angle-resolved reflectivity measurements of three structures of slightly different filling factor values. 
	The spectral window has been chosen to monitor the transformation of the same photonic band from multivalley dispersion (Fig.~\ref{fig:Experiments}c) to flatband (Fig.~\ref{fig:Experiments}d) then parabolic band (Fig.~\ref{fig:Experiments}f). 
	We observe clearly the two FW-qBICs at the valley points (Fig.~\ref{fig:Experiments}c) and their merging with the sym-BIC at the $\Gamma$ point to form a super BIC (Fig.~\ref{fig:Experiments}d). 
	All of these experimental results are in perfect agreement with the theoretical concepts shown in Figs.~\ref{fig:Concept}h-j and then numerical/analytical results shown in Figs.~\ref{fig:RealPart}a-f. 
	
	\begin{figure}[hbt!]
		\centering
		\includegraphics[width=0.48\textwidth]{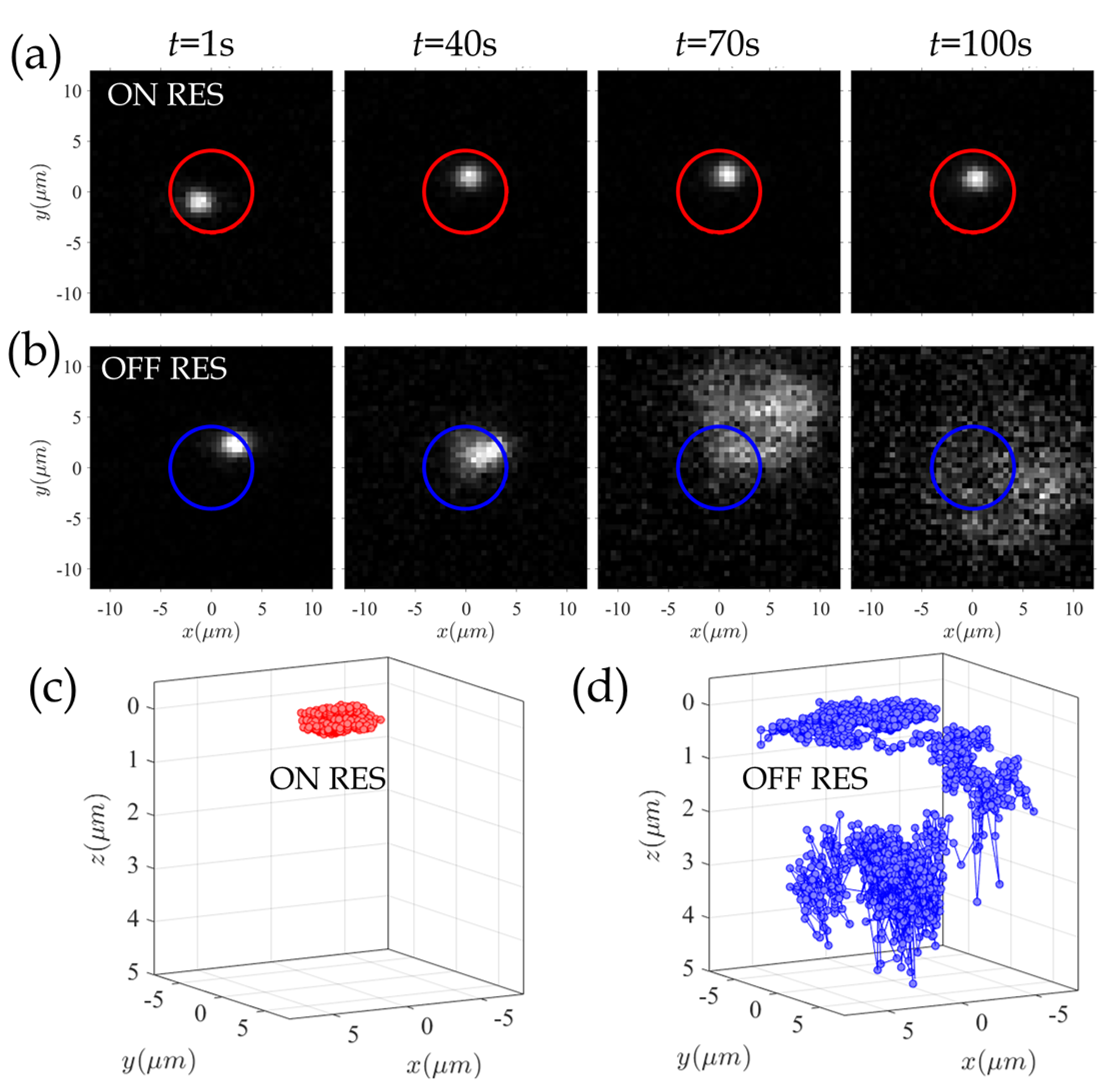} 
		\caption{(a,b) Frames captured when the laser is on resonance (a), then off resonance (b) with the super BIC. The red (blue) circle corresponds to the laser spot at on (off) resonance with the super BIC. (c,d) Coordinates of the particles extracted from the captured frames when the laser is on resonance (c), then off resonance (d) with the super BIC.}
		\label{fig:ExperimentsTrapping} 
	\end{figure}
	
	The ultraflat super BIC would find immediate use in various optoelectronic and optomechanic applications. 
	Here, as an illustration, we will employ ultraflat super BIC for the first experimental demonstration of optical trapping assisted by photonic BIC. 
	The combination of flatband and super BICs is the ideal configuration for optical trapping since: 
	i) High group index and high photon effective mass are required for a practical realization of optical trapping \cite{Fayard:22}, 
	ii) BICs exhibit strong field confinement to further improve the optical forces\cite{Yang2021,Qin:22,Hasan:23}, with the possibility to tailor novel trapping mechanisms enabled by BIC topological nature\cite{Qin:22,Qin2022}, 
	iii) Usually, nanotweezers relies on plasmonic \cite{Wong2011} or dielectric cavities \cite{Mandal2010,Milord2015,Therisod2018} that requires a nanometric position of the optical beam. 
	In contrast, the robust quality factor and energy of the ultraflat super BIC against oblique wavevectors make this photonic mode a unique means for localizing photons of long lifetime under a resonant injection pump spot without  photonic cavity. 
	Therefore, one may induce trapping hot-spots at any location of the structure via a resonant laser. Further discussions on the advantages of ultraflat super BIC for optical trapping are detailed in the Supplemental Materials\,\cite{supp}.
	In our proof-of-concept, the particles that will be trapped are polystyrene beads of \SI{1}{\micro\metre} size. 
	These particles are diluted in water ($10^7$ beads/ml) that is kept in the fluidic reservoir of an optofluidic chamber. 
	The sample, fixed to the wall of the fluidic reservoir from the inside and immersed in water, is excited by a tunable laser (1.25-\SI{1.6}{\micro\metre}) focused by a microscope objective ($\times$20, NA =0.4) to a 8.3 µm spot-size. This corresponds to a converging beam with a full angle width of approximately 10°\,\cite{supp}. 
	A fast camera is used to record a 200s movie of two sequences: the laser wavelength is set on the resonance with the ultraflat super BIC for 100s (Fig.~\ref{fig:ExperimentsTrapping}a), then set off resonance by 10 nm for the other 100s half (Fig.~\ref{fig:ExperimentsTrapping}b) (more frames are provided as videos in the Supplemental Materials\,\cite{supp}). 
	These results show that a single particle is trapped under the laser spot as soon as the laser is on resonance, and it will drift away as soon as the resonance is off. 
	From the record frames, we can extract the $x,y$ coordinates of the particle from the location of its center, and the $z$ coordinate from its apparent diameter $D$ when drifting out-of-focus, using $z=D\tan(\sin^{-1}NA)$. 
	The extracted coordinates, shown in Figs.~\ref{fig:ExperimentsTrapping}a and b, clearly demonstrate a three-dimensional trapping of single particle with ultraflat super BIC. 
	Further details and analysis on trapping mechanism with various BIC modes will be reported elsewhere.
	
	\emph{Conclusion -} 
	We introduce an original approach to engineer a new photonic state: the \textit{ultraflat super BIC}, by merging BICs located at band edges that exhibit opposite curvatures. 
	Remarkably, the ultraflat super BIC exhibits a robust high quality factor and density of states over an extensive region of the momentum space. 
	We successfully carried out the experimental demonstration using silicon gratings, where the full sequence of the merging transition was observed. 
	Furthermore, we leveraged the ultraflat super BIC state for optical trapping, demonstrating three-dimensional trapping of single particles. 
	Our results pave the way for exploring experimentally the topological forces texture associated with the topological charges of BICs \cite{Qin2022}. 
	We envision that the ultraflat super BIC holds potential for applications in low-threshold lasers \cite{Hwang2021,Ha2018} and highly sensitive sensors \cite{Yesilkoy2019,Maksimov2020,Vyas2020}.

	\begin{acknowledgments}
		This work was partly funded by the French National Research Agency (ANR) under the project CELLDance (ANR-21-CE09-0011). We thank Patrice Genevet and Qinghua Song for fruitful discussions. 
	\end{acknowledgments}

	\bibliography{Main.bib}
	




\onecolumngrid 
\newpage

\begin{center}
    \textbf{ 
    SUPPLEMENTAL INFORMATION FOR: SUPER BOUND STATES IN THE CONTINUUM ON PHOTONIC FLATBAND: CONCEPT, EXPERIMENTAL REALIZATION, AND OPTICAL TRAPPING DEMONSTRATION
    }
\end{center}



\setcounter{equation}{0}
\setcounter{figure}{0}
\setcounter{table}{0}
\setcounter{page}{1}

\renewcommand{\theequation}{S\arabic{equation}}
\renewcommand{\thefigure}{S\arabic{figure}}
\renewcommand{\vec}[1]{\boldsymbol{#1}}

\maketitle

\section{Fabrication method}
\label{sec:Fabrication} 
The fabrication starts with a \SI{2}{\micro \meter}-thick thermal oxide layer on a standard Si substrate. 
We use plasma-enhanced chemical vapor deposition (PECVD) to successively deposit three layers on this substrate: 147 nm of amorphous silicon (a-Si), 15 nm of SiO2 and 486 nm of a-Si. 
The substrate temperature is kept at 300°C for all depositions. 
The a-Si layers were obtained using SiH\textsubscript{4} as a precursor and helium as the plasma gas, the pressure in the chamber is 2 Torr and the plasma is driven by a 25W RF signal. 
The SiO\textsubscript{2} layer was fabricated by mixing SiH\textsubscript{4} and N\textsubscript{2}O (70 sccm/420sccm) at a pressure of 1 Torr and a power of 20 W. 
The 20 nm SiO\textsubscript{2} layer plays the role of an etch stop layer during the pattern transfer process to the upper a-Si layer only, to guarantee a precise control over the vertical asymmetry of the device.

The devices were written into a 300 nm-thick hydrogen silsesquioxane (HSQ) negative photoresist layer using a 30 keV e-beam lithography system. After the development in a solution of 25wt\% tetramethylammonium hydroxide (TMAH) in water, heated at 80°C, the pattern transfer into the top a-Si was carried out using inductively coupled plasma (ICP) etching with a Cl\textsubscript{2}/O\textsubscript{2} gas mixture (22 sccm/4 sccm). 
This etching recipe is highly selective and both the HSQ and SiO\textsubscript{2} stop layers remain un-etched at the end of the process. 

\begin{figure}[h]
    \centering 
    \includegraphics[width=0.5\textwidth]{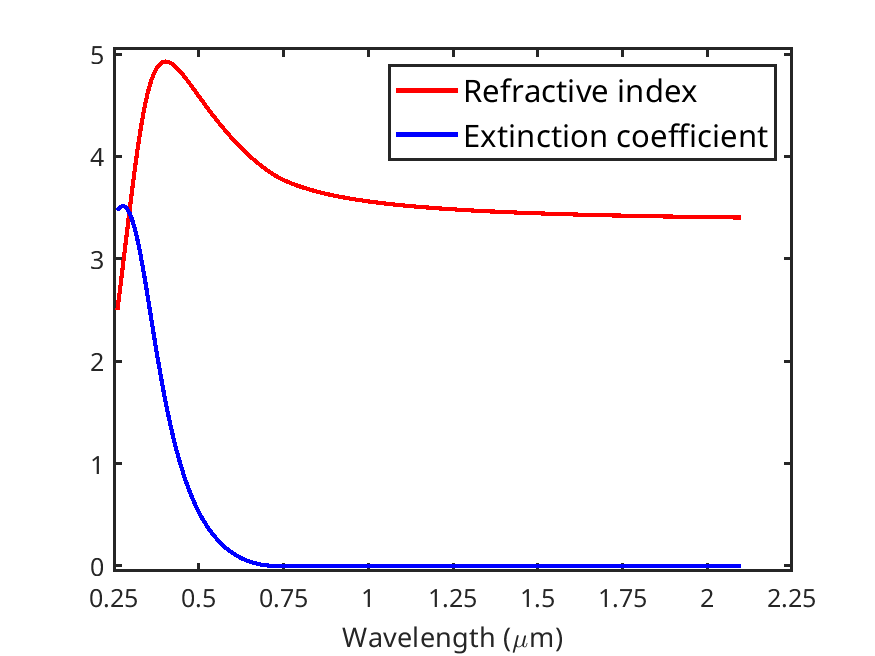} 
    \caption{The dependence of the refractive index (red) and the extinction coefficient (blue) of armorphous Si on the wavelength.} 
    \label{fig:aSi} 
\end{figure}
\begin{figure}[ht!]
\centering 
\includegraphics[width=\textwidth]{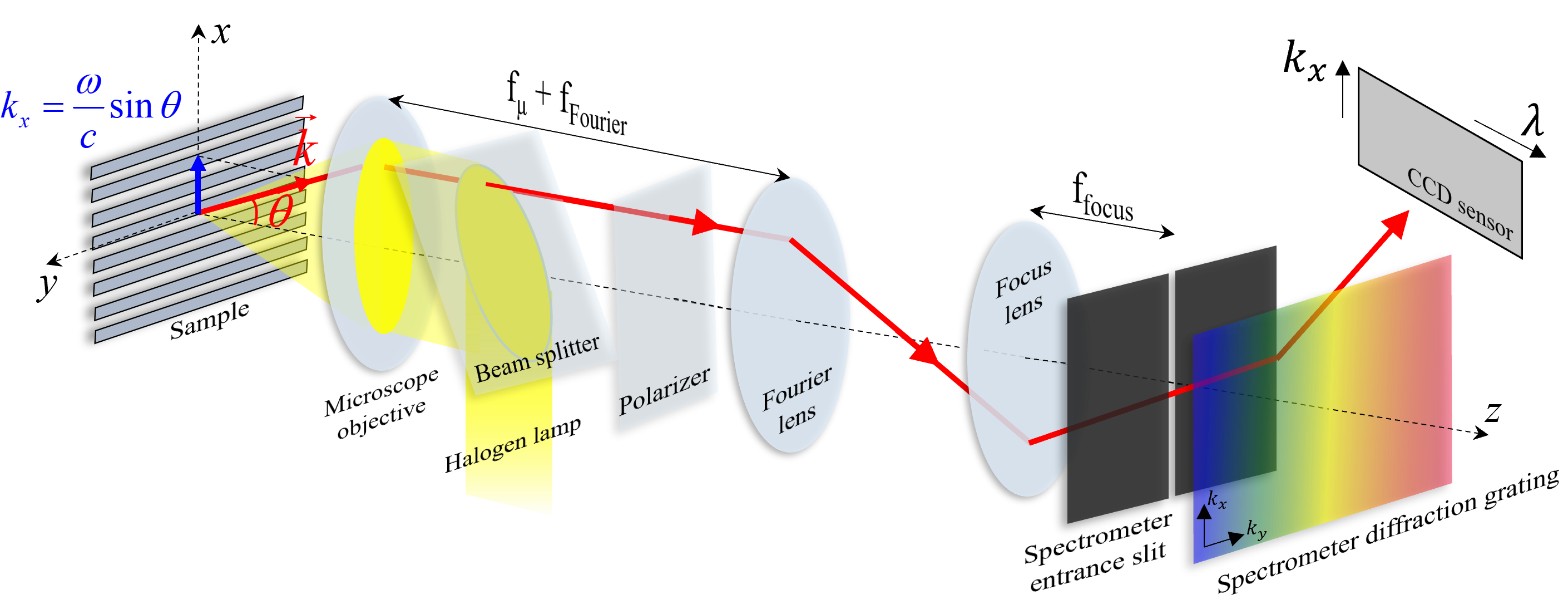}  
\caption{Home-made Fourier setup to measure the angle-resolved reflectivity of the sample in Fig.~\ref{fig:Experiments} of the main text.}
\label{fig:FourierSetup}
\end{figure}

\section{Experimental setups and results}\label{sec:ExperimentalSetup} 
\subsection{Angle-resolved reflectivity measurements}
The angle-resolved reflectivity measurements shown in Fig.~\ref{fig:Experiments} of the main text are performed by a home-made Fourier setup (see Fig.~\ref{fig:FourierSetup}). 
The sample is illuminated by a Halogen lamp and a 0.4 NA microscope objective. 
The reflection is collected by the same objective, a set of lenses, a spectrometer, and a CCD camera. 
The Fourier space of the sample, located at the Back Focal Plane (BFP) of the objective, is projected to the spectrometer entrance slit with the Fourier and the Focus lenses. 
The dispersions along the x-axis selected by the spectrometer slit is then diffracted by the spectrometer diffraction grating and projected to the CCD camera resulting in a $k_x$-$\lambda$ dispersion.

\begin{figure}[ht!]
    	\centering
    	{\includegraphics[width=1\linewidth]{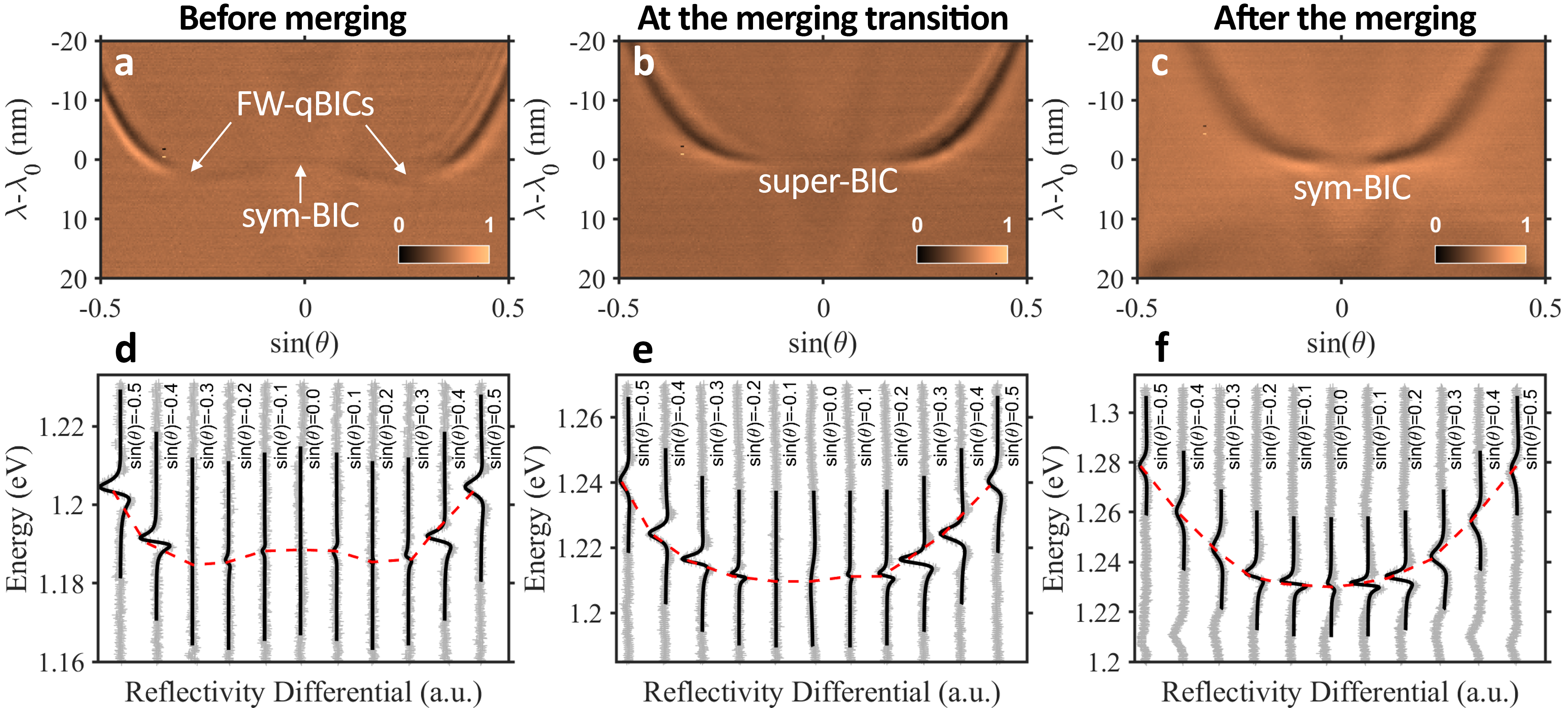}}
    	\caption{Fitting of the modes with derivative Fano resonances (d,e,f) from the cross sections at given angle $\theta$  of the derivative reflectivity $dR/d\lambda$ (a,b,c) of the measured angle-resolved reflectivity $R$ from Fig.~\ref{fig:Experiments}(c-e)of the main text.}
    	\label{fig:R1}   
    \end{figure}

\begin{figure}[ht!]
    	\centering
    	{\includegraphics[width=1\linewidth]{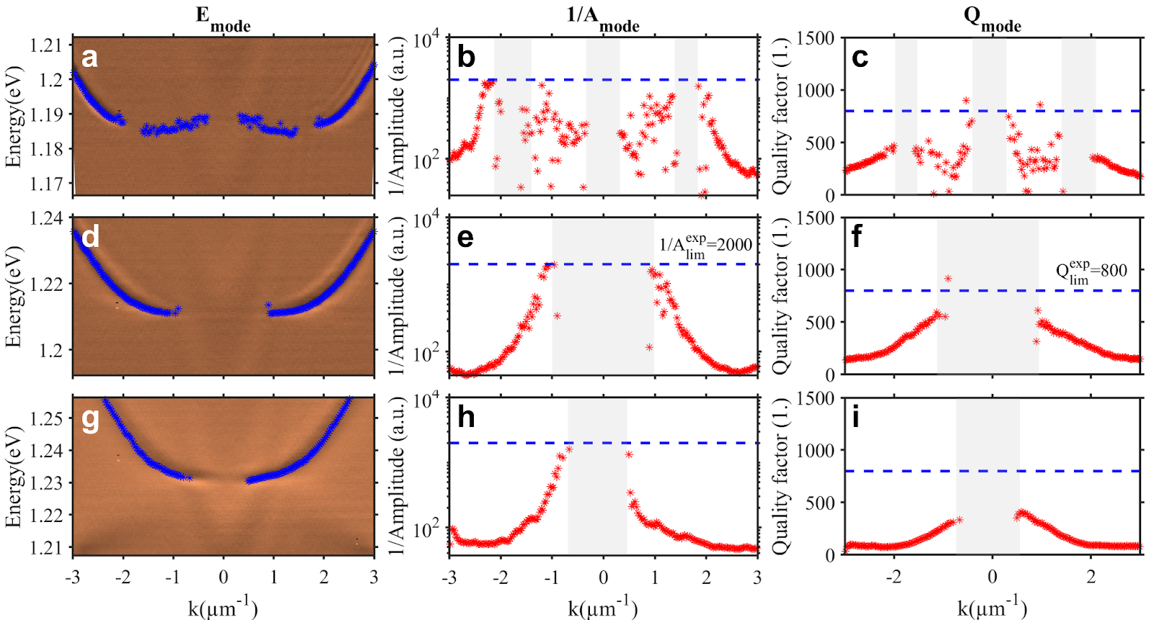}}
    	\caption{Modes energies (a,d,g), inverse amplitude (b,e,h), and quality factor (c,f,i) as a function of the wavevector from the Fano resonances fit presented in fig.\ref{fig:R1}. Only the results for which the inverse amplitude is lower than 2000 are shown, as the experimental resolution limits the quality of the fitting. In (b,e,h), the blue dash lines show the limit of 2000 of the inverse amplitude. In (c,f,i), the blue dash line represents the maximum quality factor of 800 which can be experimentally resolved.}
    	\label{fig:R2}   
    \end{figure}

 We have analyzed the Q factor of the modes shown in Fig.~\ref{fig:Experiments}(c-e) of the main text by fitting the spectra of the angle-resolved reflectivity data at a given angle $\theta$ (see Fig.~\ref{fig:R1}). 
 These configurations correspond to the dispersion before merging (i.e., multivalley), at the merging transition (ultra-flat band), and after merging (parabolic band). 
 To eliminate the parasitic reflection from the substrate, we perfomed our fitting on the derivative reflectivity $dR/d\lambda$ (see Fig.~\ref{fig:R1}a-c). 
 These mappings are much "cleaner" than the mapping of reflectivity $R$ in Fig.~\ref{fig:Experiments}(d-f).  
 We fitted the various differential reflectivity spectra at a given angle with the derivative of a Fano function (see Fig.~ \ref{fig:R1}d-f). 
 We then gathered the fitting parameters (mode energy, amplitude, and linewidth) for each spectrum, which we report in Fig. ~\ref{fig:R2}.
 

We present in Fig.~\ref{fig:R2} the mode energies (Fig.~\ref{fig:R2}a, d, g), inverse amplitudes (Fig.~\ref{fig:R2}b, e, h), and quality factors (Fig.~\ref{fig:R2}c, f, i) of the three structures studied in the article. 
Due to the spectral resolution limited to 1.5 meV from our spectrometer with a slit width of 100 µm, we could only fit Fano resonances with a maximum quality factor of 800 (due to the spectral resolution limit of our detection system), and inverse amplitude of 2000 (due to the lower signal-to-noise ratio). 
Consequently, we only show the results for modes with an inverse amplitude lower than 2000. 
We note that the measured quality factor $Q$ represents the total quality factor, including both radiative losses and other sources (in-plane losses, inhomogeneous broadening, etc.), given by $Q^{-1}=Q_{rad}^{-1} + Q_{other losses}^{-1}$. 
On the other hand, the resonant amplitude $A$ is directly related to the radiative quality factor: $A \sim Q_{rad}^{-1}$. 
Therefore, the resonant amplitude is a better metric than the total quality factor for evaluating the radiative losses of the system.

From the inverse amplitude results in Fig.~\ref{fig:R2}b, e, and h, we compare the range of wave-vectors for which we cannot resolve the modes at the BIC locations (i.e., $1/A > 2000$, with $A$ the resonant amplitude) for the three configurations. 
These regions, represented as gray areas, give an idea of the range of wave-vectors for which the radiative quality factor is extremely high and the total quality factor is higher than 500, which are respectively 0.5 and 0.66 µm$^{-1}$ for the FW-qBICs and symmetry-protected BIC in Fig.~\ref{fig:R2}b, 1.92 µm$^{-1}$ for the ultra-flat BIC in Fig.~\ref{fig:R2}e, and 1.34 µm$^{-1}$ for the symmetry-protected BIC in Fig.~\ref{fig:R2}h. 
Despite sample imperfections inducing non-radiative losses, limiting the experimental quality factor compared to theoretical predictions, our study on mode quality factors assures that the total quality factors are at least 500, and in a very large wavevector band of 1.92 µm$^{-1}$ (from -0.96 to 0.96 µm$^{-1}$) in the case of the super BIC on a flat band.

In optical trapping experiments, the focused light on the sample is a Gaussian beam with a waist of 8.3 µm, corresponding to a converging beam with a full angle width of approximately 10°. This corresponds to a range of in-plane wavevectors within approximately 1 µm$^{-1}$ (from -0.5 to 0.5 µm$^{-1}$), within which the quality factor of the BIC mode is higher than 500.

\subsection{Optical trapping measurements}
\begin{figure}[ht!]
\centering 
\includegraphics[width=0.6\textwidth]{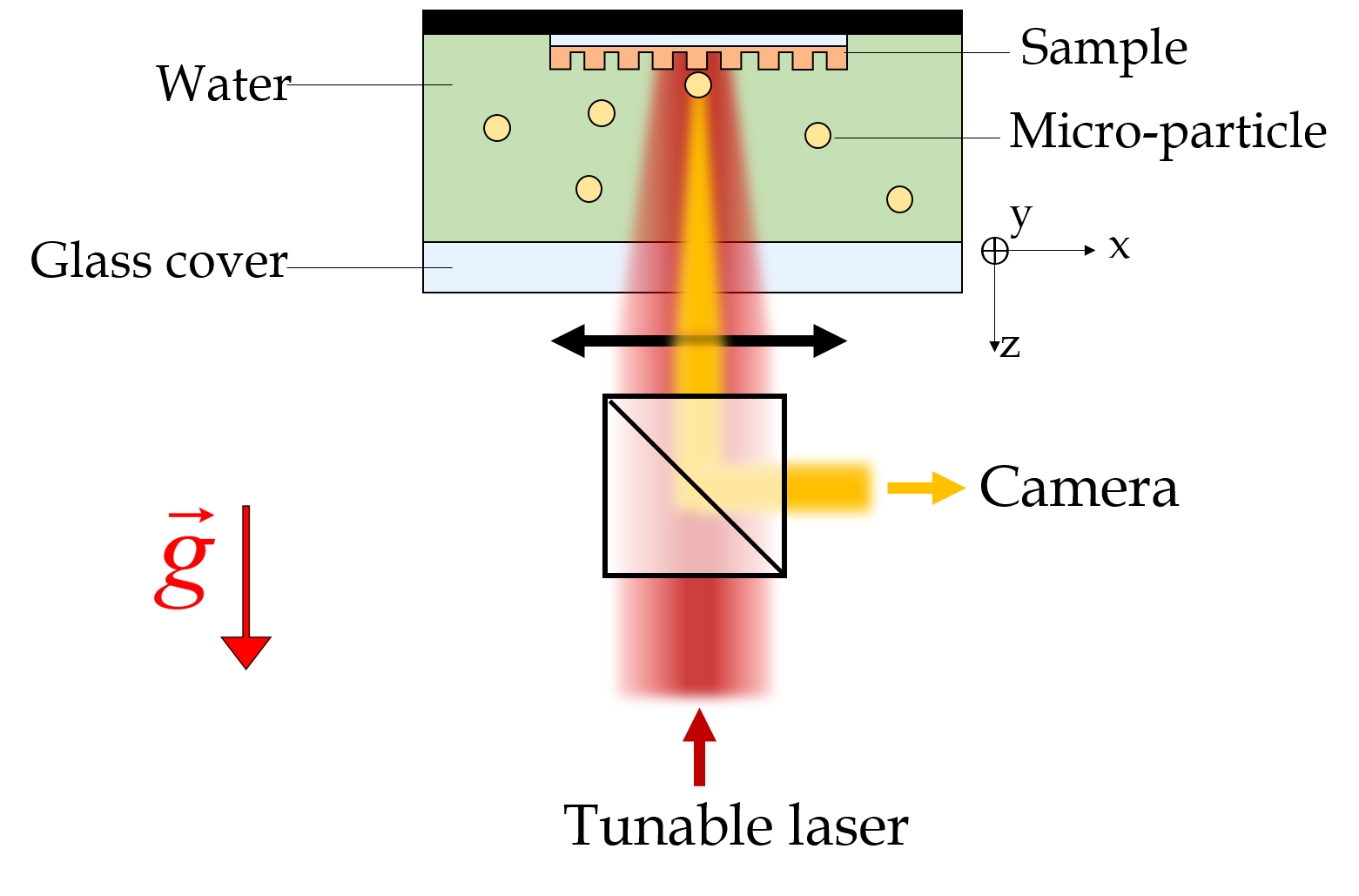}  
\caption{Home-made experimental setup for the optical trapping experiment in Fig. 5 of the main text}. 
\label{fig:OpticalTrappingSetup}
\end{figure}

\begin{figure} 
    \centering
    \includegraphics[width=\textwidth]{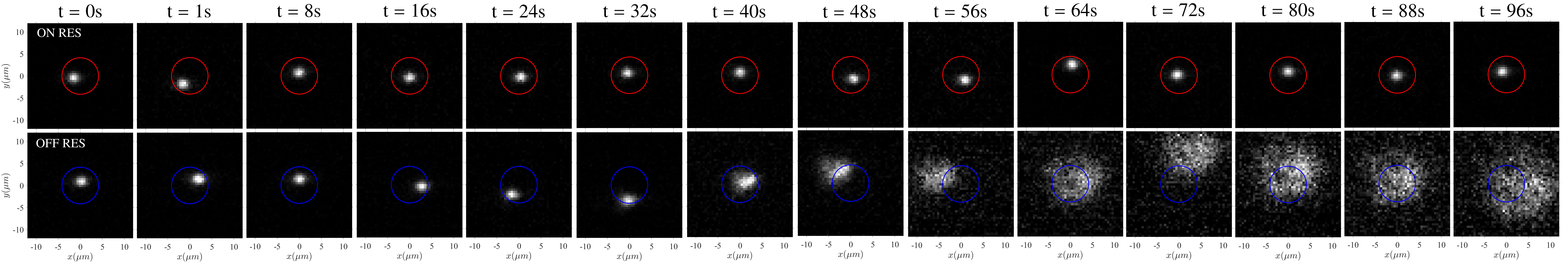}
    \caption{Frames captured when the laser is on resonance (first line) and off resonance (second line) with the ultraflat super BIC. The circles correspond to a spot of size \SI{8.3}{\micro \meter}.}  
    \label{fig:OpticalTrappingTime}
\end{figure}

The optical trapping/untrapping measurements shown in Fig.~\ref{fig:ExperimentsTrapping} of the main text are performed by a home-made setup (see Fig.~\ref{fig:OpticalTrappingSetup}). 
We use an optofluidic chamber composed of a fluidic reservoir and a substrate integrating the BIC ultra-flat. 
The fluidic reservoir allows measurements in a static liquid medium. 
The chamber is designed to allow laser excitation at the front and observations with an inverted microscope. 
Polystyrene beads of \SI{1}{\micro\metre} size are first diluted in water ($10^7$ beads / ml) in the fluidic reservoir of an optofluidic chamber.  
The sample is immersed in water and  mounted upside down when attached to the inside wall of the fluidic reservoir. Therefore, when not trapped, the beads move downward. This arrangement was chosen to prevent the beads from gravitating towards the grating.  We used a surfactant SdS (Sodium dodecyl sulfate) to avoid van der Waals interactions with the substrate and agglomeration of the beads. 
We used an optical amplifier to obtain an excitation power of a few tenths of milliwatts. 
A tunable laser (1.25-\SI{1.6}{\micro\metre}) is focused by a microscope objective (x20, NA =0.4) on the sample.  The focused light on the sample is a Gaussian beam with a waist of 8.3 µm, corresponding to a converging beam with a full angle width of approximately 10°.
The polarization of the laser beam is controlled by a polarization controller.  
The trapping/untrapping sequence is monitored by imaging the green fluorescence of polystyrene beads that are excited by a blue laser. 
The detection employs a fast camera and sensitive CMOS scientific camera with \SI{6.5}{\micro\metre} -pixels. 
The videos recorded by the camera are then analyzed to find the trajectories of the beads using Matlab.
The time evolution of the optical trapping is shown in Fig.~\ref{fig:OpticalTrappingTime}. 

\section{Destructive interference condition for Friedrich-Wintgen coupling} 
\label{sec:FWCondition}

In this section, we demonstrate the Friedrich-Wintgen condition for the destructive interference of two optical modes into a radiative channel.
The Hamiltonian of the non-hermitian system is given by: 
\begin{equation}
    H = \begin{pmatrix}
    \omega_o & U 
    \\
    U & \omega_e 
    \end{pmatrix} 
    + i 
    \begin{pmatrix}
        \gamma_o & \Gamma e^{-i\phi}  
        \\
        \Gamma e^{i \phi} & \gamma_e  
    \end{pmatrix} 
    = H_0 + i \Gamma_0   
\end{equation}

Here $\phi$ is the dephasing between the two modes into the radiative channel. 
The two matrices $H_0$ and $\Gamma_0$ are given by: 
\begin{equation} 
H_0 = 
\begin{pmatrix}
\omega_o & U + \Gamma \sin \phi 
\\
U - \Gamma \sin \phi & \omega_e  
\end{pmatrix}
, 
\Gamma_0 = 
\begin{pmatrix}
\gamma_o & \Gamma \cos \phi 
\\
\Gamma \cos \phi & \gamma_e 
\end{pmatrix}
\end{equation}

We define $\omega_0 = \dfrac{\omega_o + \omega_e}{2}$, $\Delta = \dfrac{\omega_o - \omega_e}{2}$, $\gamma_0 = \dfrac{\gamma_o + \gamma_e}{2}$, and $\delta = \dfrac{\gamma_o - \gamma_e}{2}$.  
The Hamiltonian is rewritten as follows: 
\begin{equation}
    H = 
    \begin{pmatrix}
        \omega_0 + \Delta & U + \Gamma \sin \phi \\ 
        U - \Gamma \sin \phi & \omega_0 - \Delta 
    \end{pmatrix} 
    + i 
    \begin{pmatrix}
        \gamma_0 + \delta & \Gamma \cos \phi 
        \\ 
        \Gamma \cos \phi & \gamma_0 - \delta 
    \end{pmatrix}
\end{equation}

The energy eigenvalues of the Hamiltonian have the formula:
\begin{equation}
    \Omega_{\pm} = \omega_0 + i \gamma_0 \pm 
    \sqrt{ (\Delta + i\delta)^2 + (U^2 - \Gamma^2 + i 2 U \Gamma \cos \phi ) }
\end{equation}

We define $z_1 = \Delta + i \delta$, $z_2 = a + i b$ such that $(a + ib)^2 = U^2 - \Gamma^2 + i 2 U \Gamma \cos \phi$.
We define $W = \sqrt{U^4 + 2 U^2 \Gamma^2 \cos 2 \phi + \Gamma^4}$, $V = U^2 - \Gamma^2$ and $Z = U \Gamma \cos \phi$.  
Indeed, $z_2^2$ has 2 square roots, and we can choose the one with $b \ge 0$:
\begin{equation}
    \begin{aligned}
        a =& \text{sgn} (Z) \frac{1}{\sqrt{2}} \sqrt{ W + V }
        \\ 
        b =& \frac{1}{\sqrt{2}} \sqrt{ W - V } \ge 0  
    \end{aligned} 
\end{equation}

We define $\alpha = \sqrt{z_1^2 + z_2^2}$.
The destructive configuration corresponds to minimizing the absolute value of the imaginary part of $\alpha$, leading to:
\begin{equation}
    \frac{\partial (\alpha^* - \alpha )}{\partial \Delta} = 0 
    \Leftrightarrow  
    \frac{\partial \alpha^*}{\partial \Delta} = \frac{\partial \alpha}{\partial \Delta} 
\end{equation} 

The above condition is equivalent to:
\begin{equation}
 \frac{\Delta - i \delta}{\sqrt{ (\Delta - i \delta)^2 + (a - i b )^2 }} 
 =  \frac{\Delta + i \delta}{\sqrt{ (\Delta + i \delta)^2 + (a + i b )^2 }}  
\end{equation} 

This condition implies that: 
\begin{equation}
 \frac{z_1^*}{\sqrt{ (z_1^*)^2 + (z_2^*)^2 }} = \frac{z_1}{\sqrt{z_1^2 + z_2^2}}  
 \Rightarrow 
 z_1 z_2^* \pm z_1^* z_2 = 0 
\end{equation}

We consider two cases:
\begin{enumerate}
\item \textit{Case 1:} $z_1 z_2^* + z_1^* z_2 = 0$. This case is equivalent to:
\begin{equation}
\Delta = - \delta \frac{b}{a}  
\end{equation} 

The corresponding value of $\alpha$ equals:
\begin{equation}
    \alpha_1 = \sqrt{1 - \frac{\delta^2}{a^2}} (a + ib) 
\end{equation}

Therefore, we have:
\begin{equation}
    \Omega_{1,\pm} = \left( \omega_0 \pm a \sqrt{1 - \frac{\delta^2}{a^2}} \right) + i \left( \gamma_0 \pm \sqrt{ 1 - \frac{\delta^2}{a^2} } b \right) 
\end{equation}

\item \textit{Case 2:} $z_1 z_2^* - z_1^* z_2 = 0$. This case is equivalent to:
\begin{equation}
\Delta = \delta \frac{a}{b}    
\end{equation} 

The corresponding value of $\alpha$ is equal to: 
\begin{equation}
    \alpha_2 = \sqrt{ 1 + \frac{\delta^2}{b^2}  } (a +  i b )  
\end{equation} 
\end{enumerate} 

Therefore, the energy eigenvalues are given by:
\begin{equation}
    \Omega_{2\pm} =
    \left( \omega_0 \pm a \sqrt{1 + \frac{\delta^2}{b^2}} \right) 
    + i \left( \gamma_0 \pm \sqrt{1 + \frac{\delta^2}{b^2}} b \right) 
\end{equation}  

In both cases, the mode ``$-$" has a weaker lossy rate than the mode ``$+$".
It is easy to see that $0 \le \text{Im}(\Omega_{2-}) < \text{Im}(\Omega_{1-})$. 
Therefore, the destructive interference is obtained in case 2.
The \textit{Friedrich-Wintgen condition} reads: 

\begin{equation}
    \Delta = \delta \frac{a}{b}
\end{equation} 
or it can be written equivalently under the form:

\begin{equation} 
\boxed{ 
 \omega_o - \omega_e = \text{sign} (Z) \sqrt{\dfrac{W+V}{W-V}} (\gamma_o - \gamma_e)
} 
\label{eq:FWCondition} 
\end{equation}
 
When the Friedrich-Wintgen condition \eqref{eq:FWCondition} is satisfied, the mode $\Omega_-$ is a quasi-BIC. 
The quasi-BIC becomes a BIC when $\phi = 0$ and the two polarization vectors $\vec{u}_o = \vec{u}_e$, leading to $\text{Im} (\Omega_-) = 0$.  

 
\section{Analytical theory: two-step coupling model} 
\label{sec:2StepCouplingModel} 

\subsection{General Hamiltonian} 
In this section, we present the two-step coupling model between Bloch modes in a 1D photonic crystal ``comb" structure. 
In the subsequent parts of the supplemental materials, we employ the following non-hermitian Hamiltonian, written in the basis of the guided modes $|0\rangle$ and $|1\rangle$: 
\begin{equation}
    H = d_0 \mathbb{I} + d_1 \sigma_1 + d_2 \sigma_2 + d_3 \sigma_3 
    \label{eq:GeneralHamiltonian}
\end{equation}

The eigenvalues of the Hamiltonian \eqref{eq:GeneralHamiltonian} are given by: 
\begin{equation}
    \Omega_{\pm} = d_0 \pm \sqrt{d_1^2 + d_2^2 + d_3^2}
\end{equation}
and the corresponding eigenstates are 
\begin{equation}
    |\pm \rangle = 
    \left( \begin{matrix}
    1 \\ C_{\pm} 
    \end{matrix} \right) 
\end{equation}
where 
\begin{equation}
    C_{\pm} = \frac{1}{d_1 - id_2} \left( - d_3 \pm \sqrt{d_1^2 + d_2^2 + d_3^2}  \right)    
\end{equation}

The coefficients $C_{\pm}$ allows us to calculate the farfield polarization $\vec{E}_{\pm}$ of the modes $|\pm\rangle$ from the corresponding farfield polarizations $\vec{E}_0$ and $\vec{E}_1$ of the initial modes $|0\rangle$ and $|1\rangle$ as follows: 
\begin{equation}
    \vec{E}_{\pm} = \vec{E}_0 + C_{\pm} \vec{E}_1 
\end{equation}

The unit vector $\vec{u}$ corresponding to the farfield polarization $\vec{E}$ is defined as $\vec{u} = \vec{E} / |\vec{E}|$.  

\subsection{Guided modes and polarization vectors}  
\begin{figure}
\centering 
\includegraphics[scale=0.5]{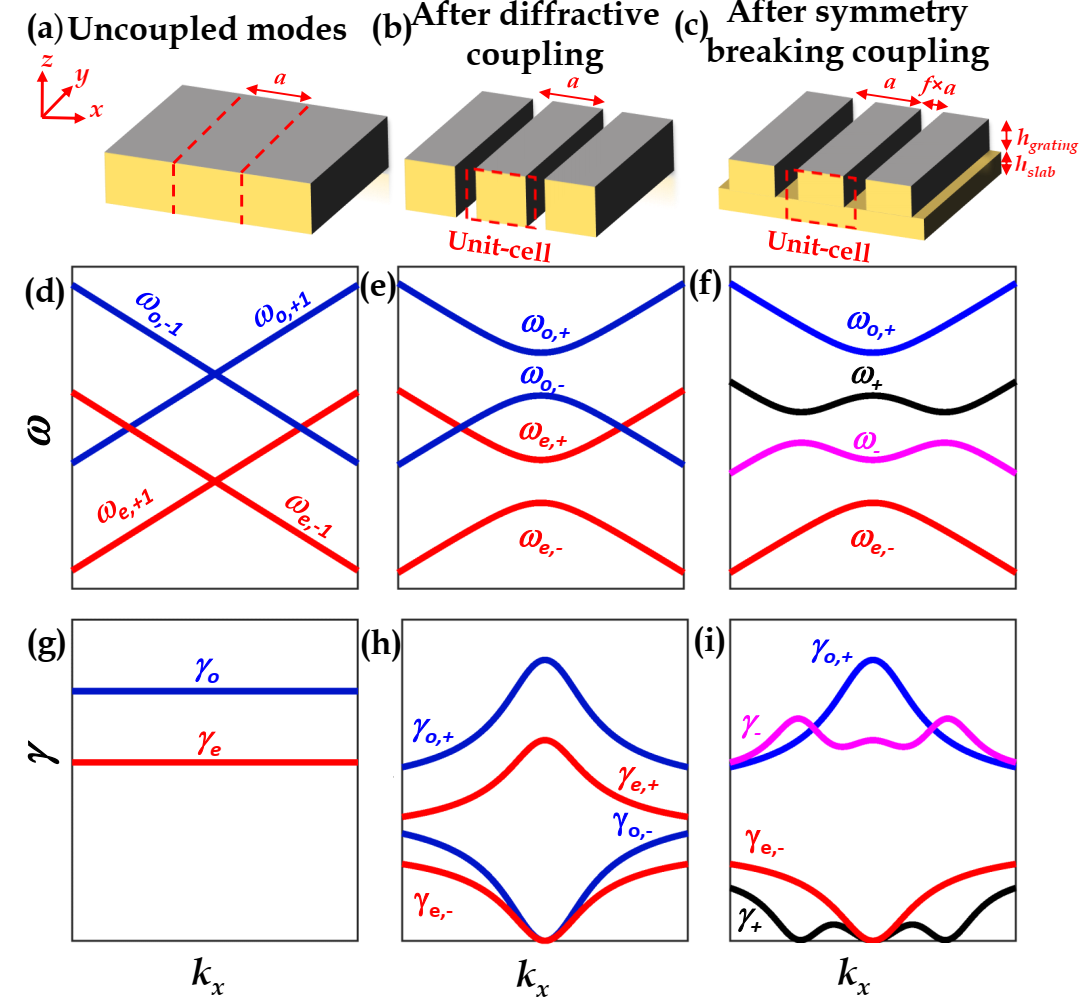}  
\caption{\textbf{Two-step coupling model.} \textit{Line 1:} (a) Homogeneous dielectric slab waveguide. (b) 1D photonic crystal with vertical symmetry. (c) 1D photonic crystal with broken vertical symmetry. \textit{Line 2:} Dispersion relations of (d) uncoupled guided modes, (e) coupled modes after diffractive coupling, and (f) coupled modes after symmetry-breaking coupling. \textit{Line 3:} Radiative loss of (g) uncoupled guided modes, (h) coupled modes after diffractive coupling, and (i) coupled modes after symmetry-breaking coupling. Here we present the dispersion curves for $k_y = 0$.}
\label{fig:TwoStepCouplingModel}
\end{figure}

We begin by considering the homogeneous dielectric slab with effective refractive index shown in Fig.~\ref{fig:TwoStepCouplingModel}a. 
Uncoupled TE guided modes in this homogeneous dielectric slab have linear dispersion (Fig.~\ref{fig:TwoStepCouplingModel}d) and constant radiative loss (Fig.~\ref{fig:TwoStepCouplingModel}g). 
In the case of interest, the fundamental mode is even, and the first excited mode is odd.   
For each parity, there are one propagating mode and a counter propagating mode.   
They have opposite group velocities and are presented as line having opposite slopes in Fig.~\ref{fig:TwoStepCouplingModel}d. 

Next, we move on the 1D photonic crystal with vertical symmetry (Fig.~\ref{fig:TwoStepCouplingModel}b).
Due to the periodicity of the photonic crystal, TE Bloch resonances can be described by crystal momenta in the Brillouin zones. 
For each parity, we describe Bloch resonances in the basis of propagating in-plane guided modes $|\vec{\beta}_m\rangle$ where $\vec{\beta}_m$ is the propagating Bloch wavevector, given by:
\begin{equation}
    \vec{\beta}_m = \left( k_x + m \frac{2\pi}{a} \right) \vec{u}_x + k_y \vec{u}_y  
\end{equation}
with $m \in \mathbb{Z}$. 

For the sake of simplicity, we define the dimensionless momentum coordinates $q_x = k_xa/(2\pi)$ and $q_y=k_ya/(2\pi)$, pulsation $\hat{\omega}=\omega a/(2\pi c)$ and velocity $\hat{v}=v/c$. 
Our analysis is restricted to the modes with $m=\pm 1$. 
That means we start by considering the four modes $\lbrace |e,\vec{\beta}_{\pm 1} \rangle, |o,\vec{\beta}_{\pm 1} \rangle \rbrace$.

In the vicinity of the $\Gamma$-point, $q_x,q_y \ll 1$, we assume that the uncoupled guided modes have linear dispersion and constant losses $\gamma_{e/o}$:  
\begin{equation}
    \begin{aligned}
    \hat{\omega}_{e,\pm 1} 
    &= \hat{\omega}_e^B + \hat{v} \left( |\beta_{\pm 1}|\frac{a}{2\pi} - 1 \right) 
    \\
    &\approx \hat{\omega}_e^B \pm \hat{v} q_x + \frac{\hat{v}}{2} q_y^2 
    \\
    \hat{\omega}_{o,\pm 1} &= \hat{\omega}_o^B + \hat{v} \left( |\beta_{\pm 1}|\frac{a}{2\pi} - 1 \right) 
    \\
    &\approx \hat{\omega}_o^B \pm \hat{v} q_x + \frac{\hat{v}}{2} q_y^2 
    \end{aligned}
\end{equation}

For those TE guided modes, the far-field polarization unit vectors satisfy $\vec{u}_{\pm 1} \cdot \vec{\beta}_{\pm 1} = 0$. 
Therefore, the polarization vectors $\vec{u}_{\pm 1}$ are given by:
\begin{equation}
    \vec{u}_{\pm 1} = \frac{1}{\sqrt{ (q_x \pm 1)^2 + q_y^2 }}  
    \left( \begin{matrix}
    -q_y
    \\
    q_x \pm 1 
    \end{matrix} \right) 
\end{equation} 

With this notation, the far-field polarization unit vectors of the uncoupled odd/even modes are: $\vec{u}_{e/o,+1} = \vec{u}_{+1} $ and $\vec{u}_{e/o,-1} = \vec{u}_{-1}$.  

Due to symmetry mismatch, the coupling between modes having opposite parities is forbidden.  
When the vertical symmetry is broken, the coupling between even-like mode and odd-like mode is weak compared to coupling between modes having the same parity. 
Therefore, instead of considering the coupling of all the four modes together, we first consider the coupling between modes having the same parity, and then consider the coupling between the resulting odd-like mode and even-like mode. 
We call the former \textit{diffractive coupling} (Fig.~\ref{fig:TwoStepCouplingModel}b,e,h) and the later \textit{symmetry-breaking coupling} (Fig.~\ref{fig:TwoStepCouplingModel}c,f,i).  
Our model is called \textit{two-step coupling model}.  
The reader can find the complete analytical model taking into account the coupling between the four modes in a single $4 \times 4$ Hamiltonian in Ref.~\cite{Letartre2022}. 

\subsection{Diffractive coupling}
The first step is diffractive coupling between modes having the same parity (even or odd).  
The Hamiltonian of the diffractive coupling is given in the basis $\lbrace |e/o,\vec{\beta}_{+1}\rangle, |e/o,\vec{\beta}_{-1}\rangle \rbrace$ by:  
\begin{equation}
    H_{e/o} = \begin{bmatrix}
    \hat{\omega}_{e/o,+1} & \kappa_{e/o}
    \\
    \kappa_{e/o} & \hat{\omega}_{e/o,-1}
    \end{bmatrix} 
    + i \begin{bmatrix}
    \gamma_{e/o} & \gamma_{e/o} (\vec{u}_{o/e,+1} \cdot \vec{u}_{o/e,-1}) 
    \\
    \gamma_{e/o} (\vec{u}_{o/e,+1} \cdot \vec{u}_{o/e,-1})  & \gamma_{e/o} 
    \end{bmatrix}
\end{equation}

By substituting 
\begin{equation}
    \begin{aligned}
    d_0 =& \frac{\hat{\omega}_{o/e,+1}+\hat{\omega}_{o/e,-1}}{2} + i \gamma_{e/o}
    \\ 
    d_1 =& \kappa_{e/o} + i \gamma_{e/o} (\vec{u}_{o/e,+1} \cdot \vec{u}_{o/e,-1})  
    \\
    d_2 =& 0 
    \\
    d_3 =& \frac{\hat{\omega}_{o/e,+1}-\hat{\omega}_{o/e,-1}}{2}
    \end{aligned}
\end{equation}
into \eqref{eq:GeneralHamiltonian},
we obtain the dispersion relation $\hat{\omega}_{o/e,\pm}$ for coupled odd and even modes.
The corresponding eigenstates $|o/e,\pm \rangle$ are given by:
\begin{equation}
    | o/e, \pm \rangle = C_{e/o,\pm} |e/o, \boldsymbol{\beta}_{+1} \rangle 
    + D_{e/o,\pm} |e/o, \boldsymbol{\beta}_{-1} \rangle  
\end{equation}
where $|C_{e/o,\pm}|^2 + |D_{e/o,\pm}|^2 = 1$. 
The corresponding far-field polarization unit vectors $\vec{u}_{o/e,\pm}$ are the linear combinations of $\vec{u}_{+1}$ and $\vec{u}_{-1}$ with the same coefficients:
\begin{equation}
    \vec{u}_{o/e,\pm} = C_{e/o,\pm} \vec{u}_{+1} 
    + D_{e/o,\pm} \vec{u}_{-1} 
\end{equation}

The dispersion and radiative loss of the coupled modes are shown in Fig.~\ref{fig:TwoStepCouplingModel}e and h, respectively. 
The diffractive coupling opens the gaps at $k_x = 0$.
The width of the gaps equals to $2\kappa_{e/o}$. 
For the radiative loss, $\gamma_{e,-}(k_x=0) = \gamma_{o,-}(k_x=0) = 0$.  
That means, for each parity, one mode gains all radiative loss, while the other has no loss, and becomes a sym-BIC at the $\Gamma$-point. 

\subsection{Symmetry-breaking coupling}
The second step is the symmetry-breaking coupling between odd-like mode $|o,+\rangle$ and even-like mode $|e,-\rangle$.  
The coupling Hamiltonian is given (in the basis of those two modes) by:
\begin{equation}
    H_b = \begin{bmatrix}
    \hat{\omega}_{o+} & U 
    \\
    U & \hat{\omega}_{e-}
    \end{bmatrix} 
    + i \begin{bmatrix}
    \gamma_{o+} & \Gamma e^{-i\phi} 
    \\
    \Gamma e^{i\phi} & \gamma_{e-} 
    \end{bmatrix} 
\end{equation}
\label{eq:symbreakHamiltonian}
where $U = \alpha |q_x|$ and $\Gamma = \sqrt{\gamma_{o+} \gamma_{e-}} (\vec{u}_{o+} \cdot \vec{u}_{e-})$.

We transform the Hamiltonian \eqref{eq:symbreakHamiltonian} into the Hamiltonian \eqref{eq:GeneralHamiltonian} by putting
\begin{equation}
    \begin{aligned}
    d_0 =& \frac{\hat{\omega}_{o+} + \hat{\omega}_{e-}}{2} + i \frac{\gamma_{o+} + \gamma_{e-}}{2} 
    \\
    d_1 =& U + i \sqrt{\gamma_{o+}\gamma_{e-}} (\vec{u}_{o+}\cdot\vec{u}_{e-}) \cos \phi  
    \\
    d_2 =&  i \sqrt{\gamma_{o+}\gamma_{e-}} (\vec{u}_{o+}\cdot\vec{u}_{e-}) \sin \phi
    \\ 
    d_3 =& \frac{\hat{\omega}_{o+} - \hat{\omega}_{e-}}{2} + i \frac{\gamma_{o+} - \gamma_{e-}}{2} 
    \end{aligned}
\end{equation}

We finally obtain the dispersion relation $\hat{\omega}_{\pm}$ and the polarization vectors $\vec{u}_{\pm}$ of the final modes $|\pm\rangle$.
The formulae of the final modes $| \pm \rangle$ are related to $|o,+\rangle$ and $|e,-\rangle$ by the relation $|\pm \rangle = C_{\pm} |o,+\rangle + D_{\pm} |e,-\rangle$ ($|C_{\pm}|^2 + |D_{\pm}|^2 = 1$).
Their polarization vectors are given by: 
\begin{equation} 
    \vec{u}_{\pm} = C_{\pm} \vec{u}_{o,+} + D_{\pm} \vec{u}_{e,-} 
\end{equation} 

The quality factor is defined from the real and imaginary parts of the modes as:
\begin{equation}
    Q_{\pm} = \frac{\text{Re} \hat{\omega}_{\pm}}{\text{Im} \hat{\omega}_{\pm}}  
\end{equation}

As shown in Fig.~\ref{fig:TwoStepCouplingModel}e, the coupling between $|o,-\rangle$ and $|e,+\rangle$ opens the gap at the crossing point, which is located at $k_x \ne 0$.  
In the vicinity of the gap, the radiative loss vanishes, resulting in a FW-qBIC which is not located at any high-symmetry point. 
By tuning the filling factor $f$, the shape of the dispersion curve changes, and this FW-qBIC moves closer to the $\Gamma$-point and merges with the sym-BIC of the mode $|e,-\rangle$ as shown in the main text.  
The quality factor calculated by the two-step coupling model is shown in Fig.~3. 
In the 2-dimensional reciprocal space, we clearly observe that the sym-BIC and the two FW-qBICs merge at the $\Gamma$-point and form a super BIC.  

\subsection{Polarization vortex and topological charge}

Knowing the electric fields $\vec{E}$ of the polarization modes, their orientation with respect to the $x$-direction is described by the polarization angle $\varphi$ \cite{Dennis2002,Saleh2007,Hecht2017}: 
\begin{equation}
    \tan 2 \varphi = \frac{2 \text{Re} ( E_x^* E_y )}{|E_x|^2 - |E_y|^2}
\end{equation}

By using this result, we calculate the topological charge around any point in the momentum space by drawing a contour $\mathcal{C}$ around this point and then calculate this formula \cite{Zhen2014}: 
\begin{equation}
    m = \frac{1}{2\pi} \oint_{\mathcal{C}} d \vec{q} \cdot \nabla_{\vec{q}} \varphi  
\end{equation}

The polarization angle $\varphi$ is calculated in the reciprocal space as shown in Fig.~3.  
These results clearly show that the topological charges of the sym-BIC and the FW-qBICs have absolute values 1. 

\subsection{Fitting the parameters} 
\label{subsec:FittingParameters} 
\begin{figure}
    \includegraphics[scale=0.6]{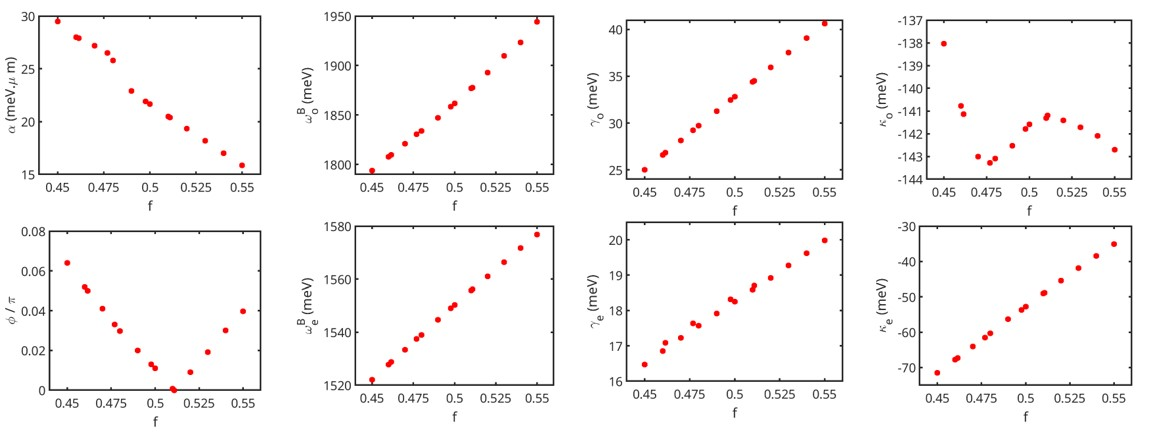}
    \caption{The parameters of the two-step coupling model as functions of the filling factor $f$}
    \label{fig:FittingParameters} 
\end{figure}

To check the validity of the two-step coupling model, we fit this analytical model with the results of the numerical RCWA method (with $k_y = 0$) to obtain the values of the parameters $\alpha$, $\phi$, $\omega_o$, $\gamma_o$, $\kappa_o$, $\omega_e$, $\gamma_e$ and $\kappa_e$ with respect to the filling factor $f$. 
The results are shown in Fig.~\ref{fig:FittingParameters}.
The group velocity for all the filling factors is $v = 0.253 c$. 
The fitting shows that $\alpha$, $\omega_o^B$, $\omega_e^B$, $\gamma_o$, $\gamma_e$ and $\kappa_e$ can be approximated as linear functions of the filling factor $f$. 
It allows us to control the coupling coefficient $\alpha$ by varying the filling factor $f$. 
The parameter $\phi$ has a minimum $\phi = 0$ at $f = 0.51$, corresponding to the destructive interference of the two modes.
The coupling strength $\kappa_o$ has a minimum at $f = 0.48$, at which the band becomes ultraflat. 

\section{Numerical results: Rigorous coupled wave analysis}

\subsection{Dispersion and Q factor}

\begin{figure}
\centering
\includegraphics[width=\textwidth]{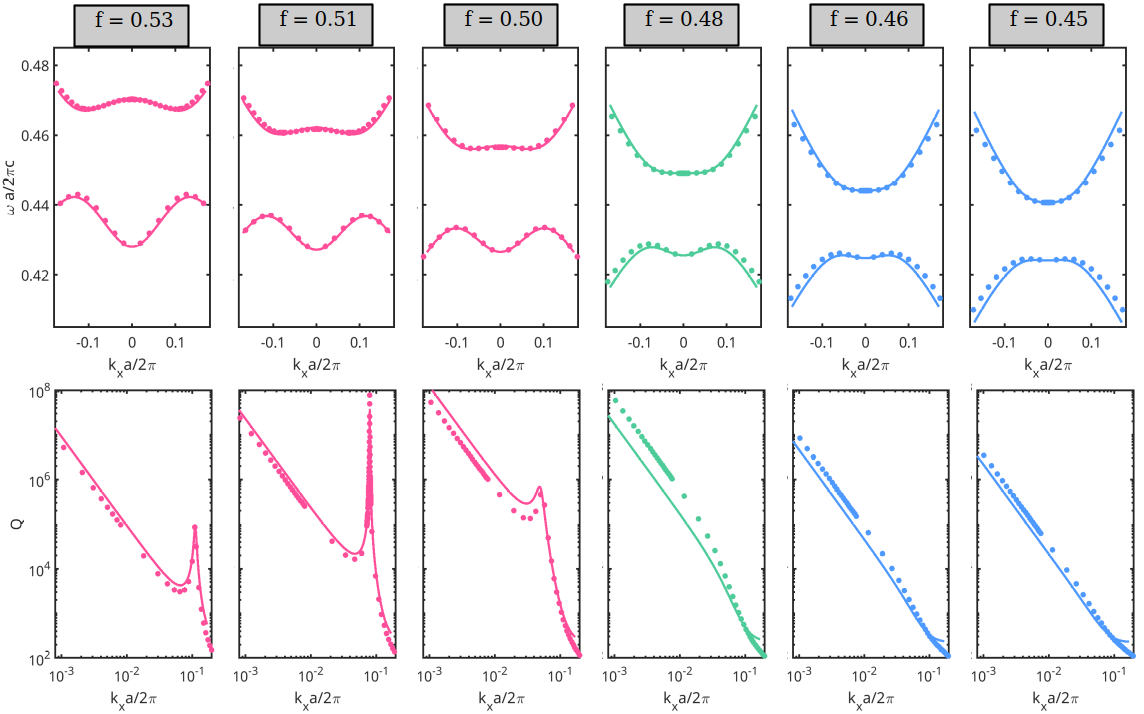} 
\caption{\textbf{Line 1: Real part of the photonic modes (dispersion relation)
}\textbf{Line 2: Imaginary part of the photonic modes (Q factor).} Dots: numerical RCWA simulations. Line: Analytical two-step coupling model. The results are calculated for $k_y = 0$.}  
\label{fig:SRCWA}   
\end{figure}

Fig.~\ref{fig:SRCWA} shows both the real part (dispersion relation) and the imaginary part (Q factor) (at $k_y = 0$) of the photon modes during the merging transition. 
The photon dispersion and Q factor are fitted from the absorption of photon modes calculated from RCWA calculations, according to the Lorentzian model \cite{Saleh2007} and are represented as the dots in the figure. 
By fitting the parameters previously shown in the last section, the analytical model perfectly reproduces both the dispersion relation and the Q factor calculated by the RCWA method, as shown by the lines in Fig.~\ref{fig:SRCWA}.  
As we vary the filling factor $f$, the dispersion relation of the upper mode transforms from multivalley (red, $f>0.48$) to ultraflat band (green, $f=0.48$) and parabola (blue, $f<0.48$).
For all structures, the Q factor tends to infinite when $k_x$ tends to 0 ($\Gamma$-point), where is located the sym-BIC.  
In the vicinity of the BIC, the Q factor obeys the law $Q \propto 1/k_x^2$ \cite{Jin2019}.  
For structures with multivalley dispersion, apart from the symmetry-protected BIC, there is a second position in the reciprocal space $(k_x=k_{BIC},k_y=0)$ where the Q factor reaches a maximum. 
This point carries another topological charge and is situated near the anticrossing point of the two bands. 
Due to its finite Q factor, it is not a true BIC but a quasi-BIC. 
For $f=0.51$, the Q factor of this state tends to infinity, corresponding to a true BIC. 
The law $Q \propto 1/|k_x-k_{BIC}|^2$ is also satisfied in the vicinity of this off-$\Gamma$ (quasi-) BIC. 
Due to the topological charge merging ($f=0.48$), there remains one BIC at the $\Gamma$-point.
In the neighborhood of the $\Gamma$-point, the Q factor satisfies $Q\propto 1/k_x^2$ instead of the $Q\propto 1/k_x^6$ law, because the off-$\Gamma$ topological charge is carried by a quasi-BIC, not by a true BIC. 
After the merging of topological charges, there remains one BIC at the $\Gamma$-point, here the $Q\propto 1/k_x^2$ law is recovered. 

\subsection{Photonic effective mass}

The photon effective mass shown in Fig.~\ref{fig:RealPart}g is calculated from the photon dispersion relation according to the following formulae:
\begin{equation}
    \frac{1}{m_x} = \frac{1}{\hbar^2} \frac{\partial^2 E}{\partial k_x^2}
\end{equation}
\begin{equation}
    \frac{1}{m_y} = \frac{1}{\hbar^2}\frac{\partial^2 E}{\partial k_y^2}
\end{equation}

In this work, we perform RCWA calculations along the two lines $k_y = 0$ and $k_x = 0$.
The corresponding components $m_x$ and $m_y$ of the photon effective mass tensor at the point $k_x = k_y = 0$ are given in Fig.~\ref{fig:RealPart}g.   
To estimate how massive a photon in our system is, we compare its effective mass with the effective mass of a photon in a Fabry-Perot cavity of half-wavelength spacer, made of the same material as our system $n_C=n_{PC}=3.15$. 
The photon has energy approximately equals to the one of photon at $\Gamma$-point (sym-BIC) $E=1700$ meV.  
Let $L$ be the thickness of the cavity, the component of the wavevector along the normal direction to the cavity is discretized $k_z L = N\pi (N \in \mathbb{Z})$. 
The phonon dispersion in this Fabry-Perot cavity is parabola in the vicinity of the point $k_{//}=0$ \cite{Lagoudakis2013}: 
\begin{equation}
    E_c (k_{//}) = E_0 + \frac{\hbar^2}{2m_c} k_{//}^2 
\end{equation}
where $E_0 = \hbar c k_z / n_C$ and the photon effective mass in the cavity $m_c = \hbar n_c k_z / c$. 
For a half-wavelength Fabry-Perot cavity ($N = 1$), the photon effective mass is $m_c/m_e=1.04\times10^{-5}$.
This value of the photon effective mass is approximately equal to the photon effective mass $m_y$ along the $y$ direction and photon effective mass $m_x$ along $x$ direction for structures having multivalley dispersion. 
For structures having parabola dispersion, the photon effective mass $m_x$ is smaller than the one in the corresponding Fabry-Perot cavity. 
For values of filling factor $f$ between the two \textit{isotropic} ($f=0.46$) and \textit{anisotropic} ($f=0.50$) configurations, the photon effective mass $m_x$ increases dramatically to infinite at ultraflat band ($f=0.48$).  

\subsection{Modification of super-BIC mode when immersed in water}
\begin{figure}[ht!]
    	\centering
    	{\includegraphics[width=0.6\linewidth]{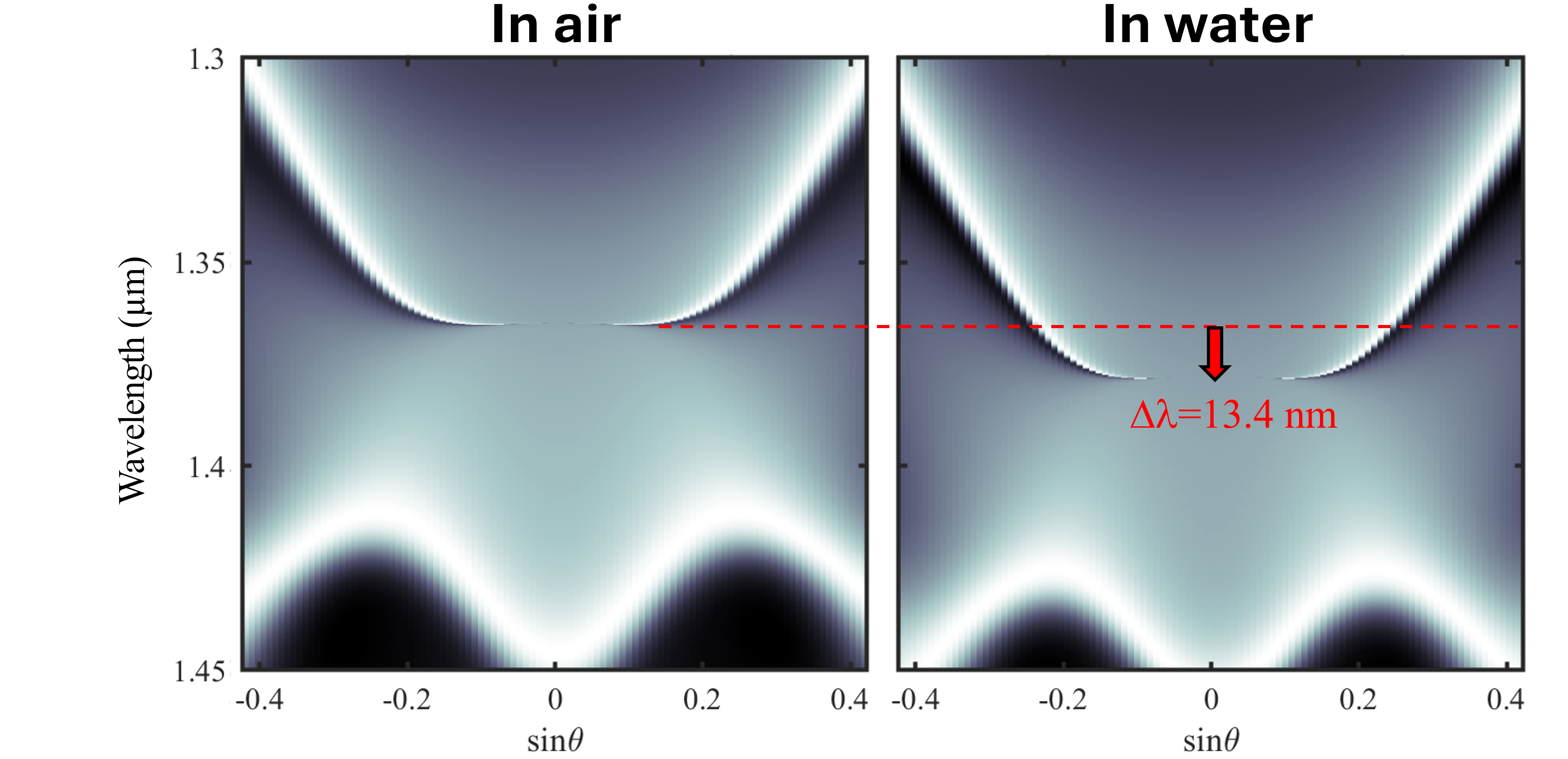}}
    	\caption{Angle-resolved reflectivity spectrum, obtained by RCWA simulations, for the same design in air (a) and in water (b).}
    	\label{fig:R3}   
    \end{figure}

To investigate the change of super-BIC on flatband mode when the air environnement is replaced by water, its dispersion in air (Fig.~\ref{fig:R3}a) and in water (Fig.~\ref{fig:R3}b) are compared. 
The results, as shown in Fig.~\ref{fig:R3}, confirm that the flatband remains robust when the sample is submerged in water, experiencing only a red shift of 13.4 µm. 
We anticipated this shift in the trapping experiment; hence, we systematically conducted a reflectivity measurement in water to determine the resonance wavelength before the trapping experiment.

\subsection{Figure of merit to quantify quality factor robustness and band flatness}
The super-BIC on flatband merges the engineering of the imaginary part of the photonic band (by merging BICs) with the real part (by creating a flatband). 
This unique approach results in a photonic species that demonstrates: 
i/ a robustly high quality factor at oblique angles due to the super BIC, 
and ii/ an extremely high density of states with resonance energy independent of the angle due to the flatband. 
This combination leads to new applications that neither super BIC states nor flatband states alone could achieve.

\begin{figure}[ht!]
\centering
{\includegraphics[width=0.65\linewidth]{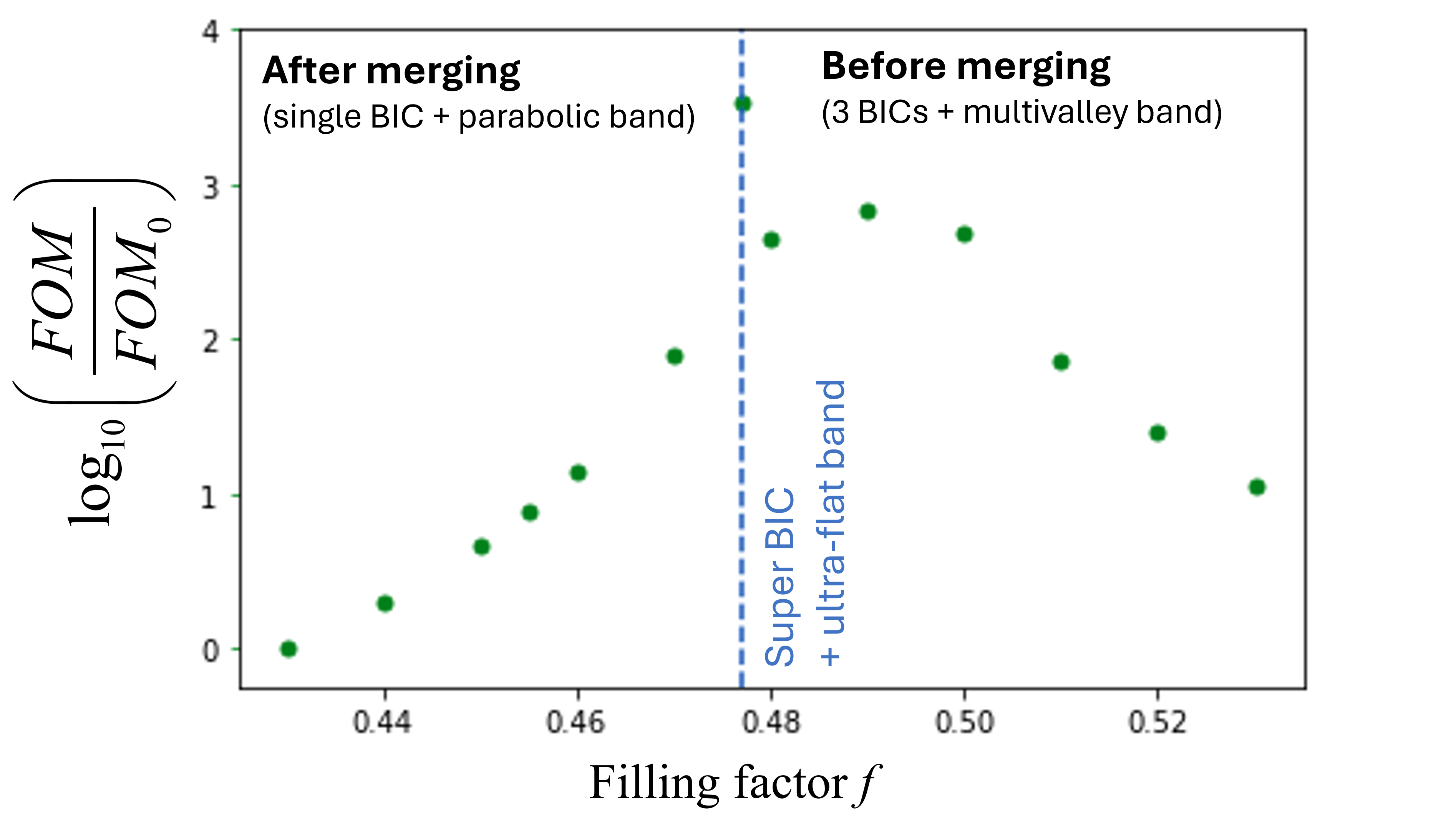}}
\caption{The figure of merit $FOM$, as defined by Equation \ref{eq:FOM}, for different values of the filling factor $f$.}
\label{fig:R6}
\end{figure}

To provide further clarity, we introduce a figure of merit (FOM) that quantifies both the robustness of the quality factor $Q$ at oblique angles and the flatness of the photonic band. 
The FOM is defined as:
\begin{equation}
FOM = \frac{m}{\langle\gamma\rangle}\label{eq:FOM}
\end{equation}
where $\langle\gamma\rangle$ is the average of the mode linewidth $\gamma$ for incident angles from 0 to 10°, and $m$ is the photon effective mass along the $x$ direction. 
Figure \ref{fig:R6} shows the FOM at various filling factors $f$, with values normalized to $FOM_0$, the FOM at a filling factor of $f=0.43$. 
These results clearly illustrate a several-orders-of-magnitude enhancement of the FOM for the super BIC on a flatband.

In reinforcing our argument, an enhancement of the optical trapping mechanism of several orders of magnitude can be obtained when combining the properties of the super BIC and flatband (see section \ref{sec:trappingperformance}. 

\section{Numerical simulations: Finite Elements Method}  
\subsection{Topological transition}
\begin{figure}[t] 
\centering 
\includegraphics[scale=0.4] {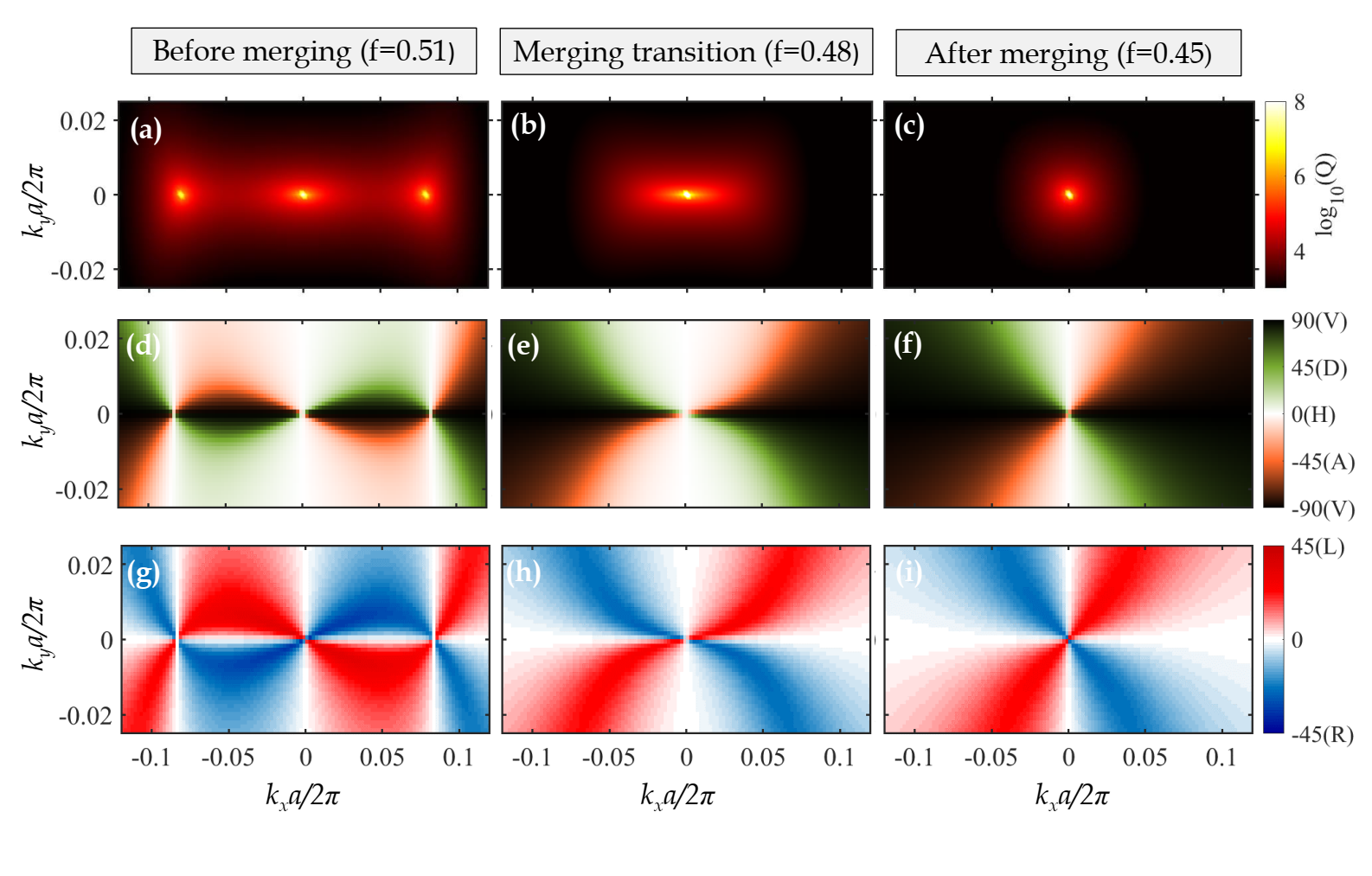}  
\caption{\textbf{Numerical simulations using the Finite Element Method.} \textit{Line 1:} Quality factor of modes in the 2-dimensional reciprocal space before merging (a), at merging transition (b), and after merging (c). \textit{Line 2:} The polarization angle $\phi$ in the 2-dimensional reciprocal space before merging (d), at merging transition (e), and after merging transition (f). \textit{Line 3:} The ellipticity $\chi$ in the 2-dimensional reciprocal space before merging (g), at merging transition (h), and after merging transition (i). The results are calculated by using the COMSOL software.}
\label{fig:COMSOLReciprocalSpace}
\end{figure}

\begin{figure}
    \centering
    \includegraphics[scale=0.55]{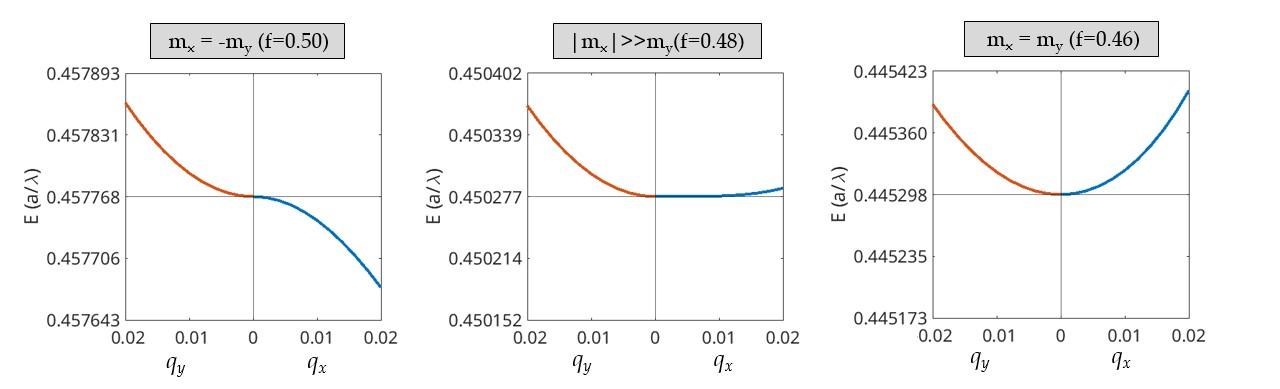} 
    \caption{The dispersion curves with $q_x = 0$ (red), $q_y = 0$ (blue) for the filling factors $f = 0.50, 0.48, 0.46$. Here $q_x = k_xa/(2\pi)$ and $q_y = k_ya/(2\pi)$. The three cases have $m_x = m_y$, $|m_x| \gg m_y$ and $m_x = -m_y$, respectively.} 
    \label{fig:MxMyTransform}
\end{figure}

The results of the two-step coupling model are perfectly reproduced by the numerical Finite Elements Method using the COMSOL software. 
For filling factor $f=0.51$, numerical results (Fig.~\ref{fig:COMSOLReciprocalSpace}a) show one sym-BIC and two FW-qBICs located at the same situations in the reciprocal space as the analytical two-step coupling model (Fig.~\ref{fig:MergingTransition}).  
The merging transition is also observed for filling factor $f=0.48$ (Fig.~\ref{fig:COMSOLReciprocalSpace}b), and a sym-BIC remains robust at filling factor $f=0.45$. 
The corresponding maps of the polarization angle $\varphi$ (Fig.~\ref{fig:COMSOLReciprocalSpace}c,d,e) absolutely agree with the ones calculated by the two-step coupling model, showing that the sym-BIC has topological charge -1, the FW-qBICs have topological charge +1, and the total topological charge is conserved at the merging transition. 
We also present the map of the ellipticity angle $\chi$ in Fig.~\ref{fig:COMSOLReciprocalSpace}g,h,i.     
The dispersion surfaces of the isotropic configurations are shown in Fig.~2h.   
We also visualize the dispersion curves along the lines $k_x = 0$ (red) and $k_y = 0$ (blue) of the cases $m_x = m_y$, $|m_x| \gg m_y$ and $m_x = - m_y$ in Fig.~\ref{fig:MxMyTransform}. 

\subsection{Discussion on evanescent field and optical trapping performance}\label{sec:trappingperformance}

\begin{figure}[ht!]
    	\centering
    	{\includegraphics[width=0.8\linewidth]{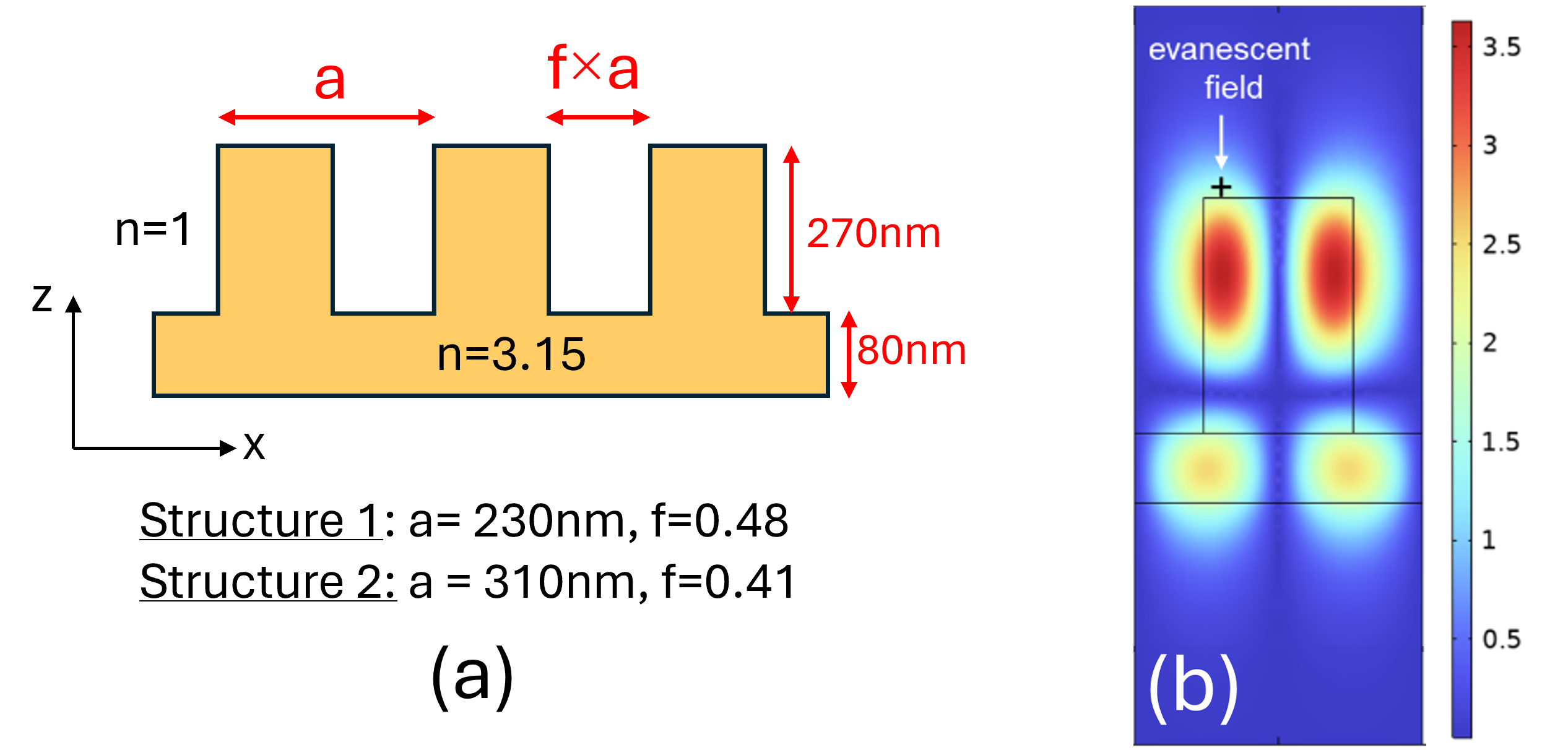}}
    	\caption{(a) Two designs to evaluate the trapping performances. (b) The field distribution of the super BIC resonance, showing important evanescent field. This result is obtained by eigenmode simulations via COMSOL.}
    	\label{fig:R4}   
    \end{figure}

To discuss the role of super-BIC and flatband for optical trapping performance, simulations using the finite element method have been conducted on two structures (Fig.~\ref{fig:R4}a). 
One corresponds to a super BIC on flatband dispersion (at the merging transition), and the other to a single BIC on parabolic dispersion (after the merging). 
The design parameters were chosen so that the two resonances have the same energy at the $\Gamma$ point. 
As discussed below, our results show not only a significant enhancement of the near field in both cases but also the superior performance of super BIC on flat dispersion compared to single BIC on parabolic ones.

\begin{figure}[ht!]
    	\centering
    	{\includegraphics[width=1\linewidth]{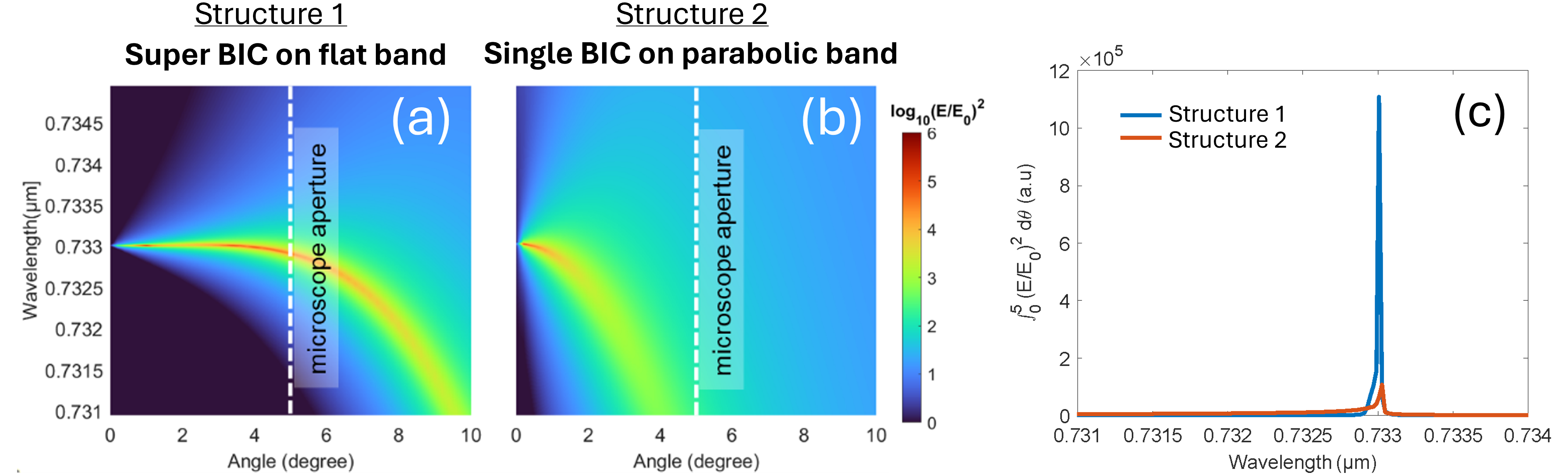}}
    	\caption{(a,b) Enhancement of the evanescent field when scanning incident angle and wavelength. Here the evanescent field is monitored at the cross point in Fig.\ref{fig:R4}b. (c) The enhancement factor when integrated from 0° to 5°.}
    	\label{fig:R5}   
    \end{figure}

The field distribution of the super BIC eigen-mode is shown in Fig.~\ref{fig:R4}b, showing clearly a strong evanescent field outside the grating. 
We then simulated the optical response of the structure when excited by plane waves, with the incident angle $\theta$ ranging from 0° to 10°. 
The cross in Fig.~\ref{fig:R4}b indicates the position where the value of the evanescent field is monitored while scanning both the incidence angle $\theta$ and wavelength $\lambda$. 
The results are depicted in Fig.~\ref{fig:R5}a,b, showing $(E/E_0)^2$ (on a log scale) versus $\lambda$ and $\theta$. 
Here, $E$ is the value of the electric field (at the cross position in the evanescent field) and $E_0$ is the value of the electric field of the incident beam. 
We note that optical forces are proportional to the square of the electric field, thus the mappings in Fig.~\ref{fig:R5} correspond to the optical force when scanning the incident angle and wavelength.

We observed that in both cases, an enormous enhancement of up to six orders of magnitude is seen for both structures. 
However, there are two striking differences between these structures. 
First, the enhancement reduces quickly at oblique angles for the single BIC configuration (Fig.~\ref{fig:R5}b) due to the rapid decrease of the quality factor, but it remains very robust at oblique angles for the super BIC configuration (Fig.~\ref{fig:R5}a), thanks to the extended high-quality factor. 
Second, the wavelength of the enhancement resonance depends strongly on the incident angle in the case of the parabolic band (Fig.~\ref{fig:R5}b) but remains constant for incident angles from 0° to 5° in the case of the flatband (Fig.~\ref{fig:R5}a). 
These two features lead to the superiority of the super BIC on flatband in trapping performance. 
Indeed, in the trapping experiment, we use a microscope objective to excite the structure with incident angles ranging from 0° to 5°. 
Therefore, we can integrate the value of $(E/E0)^2$ over angles from 0° to 5° for each excitation wavelength as a first simple approach to assess the influence of the finite aperture of the experimental setup. 
The results, depicted in Fig.~\ref{fig:R5}c, clearly show that the field enhancement at resonance is almost 20 times larger for the ultraflat dispersion structure compared to the parabolic ones. 
This final result clearly demonstrates the superiority of super BIC on flatband over single BIC on parabolic band for optical trapping.



	
\end{document}